\documentclass{jfm}

\usepackage{graphicx}
\usepackage{natbib}

\usepackage{latexsym}
\usepackage{amsmath}
\usepackage{amssymb}
\usepackage{dcolumn}% Align table columns on decimal point
\usepackage{bm}% bold math
\usepackage{wasysym}
\usepackage{mathtools}
\usepackage{color}
\usepackage{MnSymbol}
\usepackage{comment}
\ifCUPmtlplainloaded \else
  \checkfont{eurm10}
  \iffontfound
    \IfFileExists{upmath.sty}
      {\typeout{^^JFound AMS Euler Roman fonts on the system,
                   using the 'upmath' package.^^J}%
       \usepackage{upmath}}
      {\typeout{^^JFound AMS Euler Roman fonts on the system, but you
                   dont seem to have the}%
       \typeout{'upmath' package installed. JFM.cls can take advantage
                 of these fonts,^^Jif you use 'upmath' package.^^J}%
      }
  \else
  \fi
\fi

\ifCUPmtlplainloaded \else
  \checkfont{msam10}
  \iffontfound
    \IfFileExists{amssymb.sty}
      {\typeout{^^JFound AMS Symbol fonts on the system, using the
                'amssymb' package.^^J}%
       \usepackage{amssymb}%

      }{}
  \fi
\fi

\ifCUPmtlplainloaded \else
  \IfFileExists{amsbsy.sty}
    {\typeout{^^JFound the 'amsbsy' package on the system, using it.^^J}%
     \usepackage{amsbsy}}
    {\providecommand\boldsymbol[1]{\mbox{\boldmath $##1$}}}
\fi

\newsavebox{\astrutbox}
\sbox{\astrutbox}{\rule[-5pt]{0pt}{20pt}}

\newcommand{\argch}{\mathrm{argch}} % argument cosinus hyperbolique
\newcommand{\degree}{\ensuremath{^\circ}}

\title[Drop impact entrapment of bubble rings]{Drop impact entrapment of bubble rings}

\author[M.-J. Thoraval, K. Takehara, T. G. Etoh and S. T. Thoroddsen]
{M.-J.\ns T\ls H\ls O\ls R\ls A\ls V\ls A\ls L$^1$,
K.\ns T\ls A\ls K\ls E\ls H\ls A\ls R\ls A$^2$,\vspace{0.08 in}\break \ns
T.\ns G.\ns E\ls T\ls O\ls H$^2$ \and 
S.\ns T.\ns T\ls H\ls O\ls R\ls O\ls D\ls D\ls S\ls E\ls N$^1$\thanks{Email address for correspondence: 
sigurdur.thoroddsen@kaust.edu.sa}}

\affiliation{$^1$Division of Physical Sciences and Engineering and Clean Combustion Research Center,
King Abdullah University of Science and Technology (KAUST), Thuwal 23955-6900, Saudi Arabia\\[\affilskip]
$^2$Department of Civil and Environmental Engineering, Kinki University, Higashi-Osaka 577-8502, Japan}

\pubyear{2012}
\volume{???}
\pagerange{???--???}
\date{7 August 2012; revised 5 November 2012} %; revised ?; accepted ?. - To be entered by editorial office}
\begin{document}

\maketitle

%\preprint{APS/123-QED}
\begin{abstract}
We use ultra-high-speed video imaging to look at the initial contact of a drop impacting onto a liquid layer.
We observe experimentally the vortex street and the bubble-ring entrapments predicted numerically, 
for high impact velocities, by Thoraval {\it et al.} [{\em Phys.\ Rev.\ Lett.\/} \textbf{108}, 264506 (2012)].
These dynamics occur mostly within 50 $\mu$s after the first contact, requiring imaging at 1 million frames/sec.
For a water drop impacting onto a thin layer of water,
the entrapment of isolated bubbles starts through azimuthal instability, 
which forms at low impact velocities, in the neck connecting the drop and pool.
For $Re$ above about 12\,000, up to 10 partial bubble-rings have been observed at the base of the ejecta,   
starting when the contact is $\sim$ 20\% of the drop size.
More regular bubble rings are observed for a pool of ethanol or methanol.
The video imaging shows rotation around some of these air cylinders,
which can temporarily delay their breakup into micro-bubbles.
The different refractive index in the pool liquid reveals the destabilization of
the vortices and the formation of streamwise vortices and intricate vortex tangles.
Fine-scale axisymmetry is thereby destroyed.
We show also that the shape of the drop has a strong influence on these dynamics.
\end{abstract}

\section{Introduction}
\label{sec:Intro}
The impact of a drop onto a pool surface has been studied for over a century,
but revolutionary improvements in high-speed video technology \citep{Etoh2003}
have recently opened up this canonical geometry to renewed scrutiny.
This applies especially to the earliest contact between the drop and pool,
where intricate details have emerged and play a crucial role during air entrapment and splashing
\citep{Yarin2006, Thoroddsen2008}. 

The impact of a drop always entraps a bubble under the center of the drop,
as a disk of air is produced by the lubrication pressure and rapidly contracts into a bubble at the center
\citep{Thoroddsen2003, LiowCole2007, Korobkin2008, Mani2010, HicksPurvis2010, DriscollNagel2011, Kolinski2012, vanderVeen2012}.
Following this central air disk entrapment on a liquid pool, the outer contact forms a neck,
which emits an ejecta sheet for sufficiently large Reynolds numbers
\citep{Thoroddsen2002, WeissYarin1999, Davidson2002, JosserandZaleski2003, Howison2005}.
These ejecta are the source of the finest spray droplets \citep{Thoroddsen2011, Zhang2012},
which is of relevance to numerous processes, such as combustion and aerosol formation.

However, at even larger impact energy, these ejecta give way to random splashing 
of small droplets, see \cite{Thoroddsen2002}.
Numerical simulations by \citet{Thoraval2012} have shown that the base of the ejecta can become unstable,
bending up and down as the free surface sheds alternate sign vortex rings into the liquid 
and often entrapping bubble rings.
These bubble rings alternate between the top and bottom sides of the ejecta.
This regime is the focus of the current investigation.

Very recent experiments \citep{CastrejonPita2012} have used side-view and laser-induced fluorescence 
to verify the presence of the von K{\'a}rm{\'a}n street for conditions similar to those in \citet{Thoraval2012}.
Herein we show the first experimental observations of the formation of the bubble tori.

The main axisymmetric features of the vortex street and bubble rings entrapments
are observed experimentally.
However, three-dimensional effects rapidly break the symmetry.
Herein, we show that even at rather modest impact velocities,
azimuthal instabilities can appear in the neck between the drop and the pool.
Imaging using two different liquids also reveals the shedding of streamwise vortices
and their intricate dynamics, similar to three-dimensional instabilities
of the cylinder wake \citep{Williamson1996}, or the shear layer \citep{LasherasChoi1988}.
These intricate structures have perhaps escaped earlier experimental notice as they 
develop in a sub-millimeter region and evolve in less than 50~$\mu$s.

\begin{figure}
\begin{center}
\vspace{0.1 in}
\includegraphics[width=0.88\linewidth]{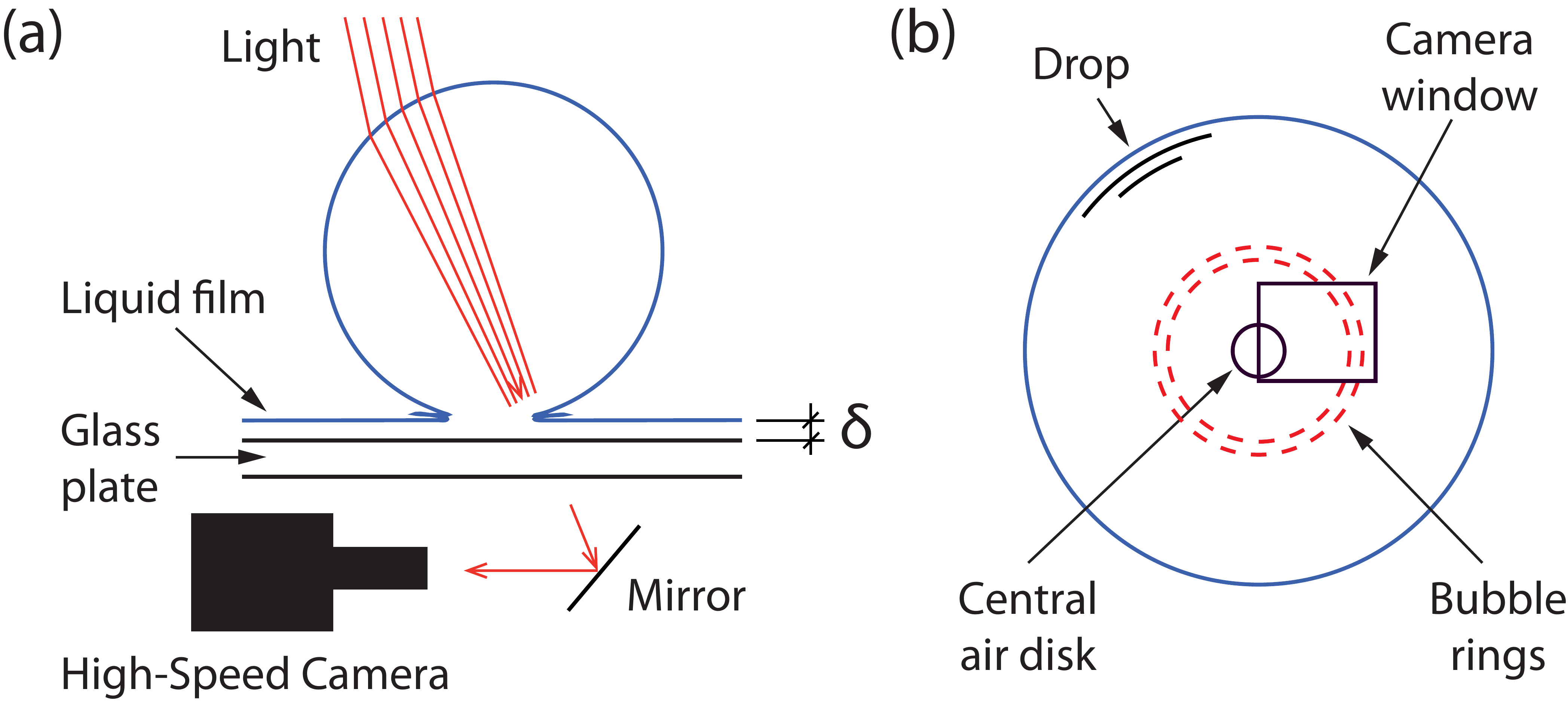}\vspace{0.1 in}\\
\end{center}
\caption{(a) Imaging setup. We use backlight imaging of the
drop impact from below, through a glass plate.
The drop acts as a lens focusing the illumination
to a limited area on the observation window.
Several lights were used in some cases to obtain a larger illuminated area.
(b) Camera viewing area shown in bottom view,
corresponding to the area in Fig.~\ref{Fig_06}(c).
}
\label{Fig_01}
\end{figure}

\section{Experimental Setup and Numerics}
\label{sec:ExperimentsAndNumerics}
\subsection{High-Speed Video Imaging}
\label{sec:Experiments}
In this work we image drop impacts onto shallow pools through a bottom glass plate (Fig.~\ref{Fig_01}).
Limited imaging (only in Fig. 16(a,b)) was done from the side above the pool surface. 
We use identical water drops in the entire study, while changing the composition of the pool liquid.
The pool liquids tested are water, ethanol and methanol which are all highly miscible with the water drop.
The liquid properties are given in Table \ref{table01}.
The difference in refractive index between the water drop and the ethanol or methanol pools allows us to image
the flow structures as they distort the interface between the two liquids.

The use of shallow pools, or thin films, is dictated by the need to change the pool liquid following
every impact as well as by the optical setup, where the limited focal distance of the long-distance
microscope rules out bottom views through deep pools.

The pool depths $\delta$ were varied from about 25 $\mu$m to 1 mm.
Here we use a long-distance microscope for magnifications up to about 15 % 9.2 
for maximum pixel resolution of $\sim 4.1 \; \mu$m/px,
when using the Shimadzu Hypervision CCD video camera \citep{Etoh2003}, at frame rates up to 1 million fps.
Some of the imaging was also done at a lower frame rate with a Photron SA5 CMOS camera,
with a magnification up to 10 and maximum pixel resolution $\sim 2 \; \mu$m/px.
Using thin bottom layers restricts the vertical motion of the interface
between the drop and the pool liquid during the impact,
thereby making well-focused imaging easier with the limited focal distance.
For further optical/triggering details see \citet{Thoroddsen2012}.

The drop is pinched from a 3 mm nozzle, to produce an effective drop diameter of $D = (D_v D_h^2)^{1/3} = 4.67$ mm,
where $D_v$ and $D_h$ are the instantaneous vertical and horizontal diameters.
We characterize the impact conditions by the Reynolds number $Re$,
the Weber number $We$ and the splashing parameter $K$, defined as:
$$Re = \frac{\rho DV}{\mu}, \;\;\;\;\;\;\;\; We = \frac{\rho DV^{2}}{\sigma}, \;\;\;\;\;\;\;\; K = We \sqrt{Re},$$
where $\rho$, $\mu$ and $\sigma$ are respectively the density, dynamic viscosity and surface tension of the drop liquid,
and $V$ the drop impact velocity.

\begin{figure}
\begin{center}
\includegraphics[width=0.93\linewidth]{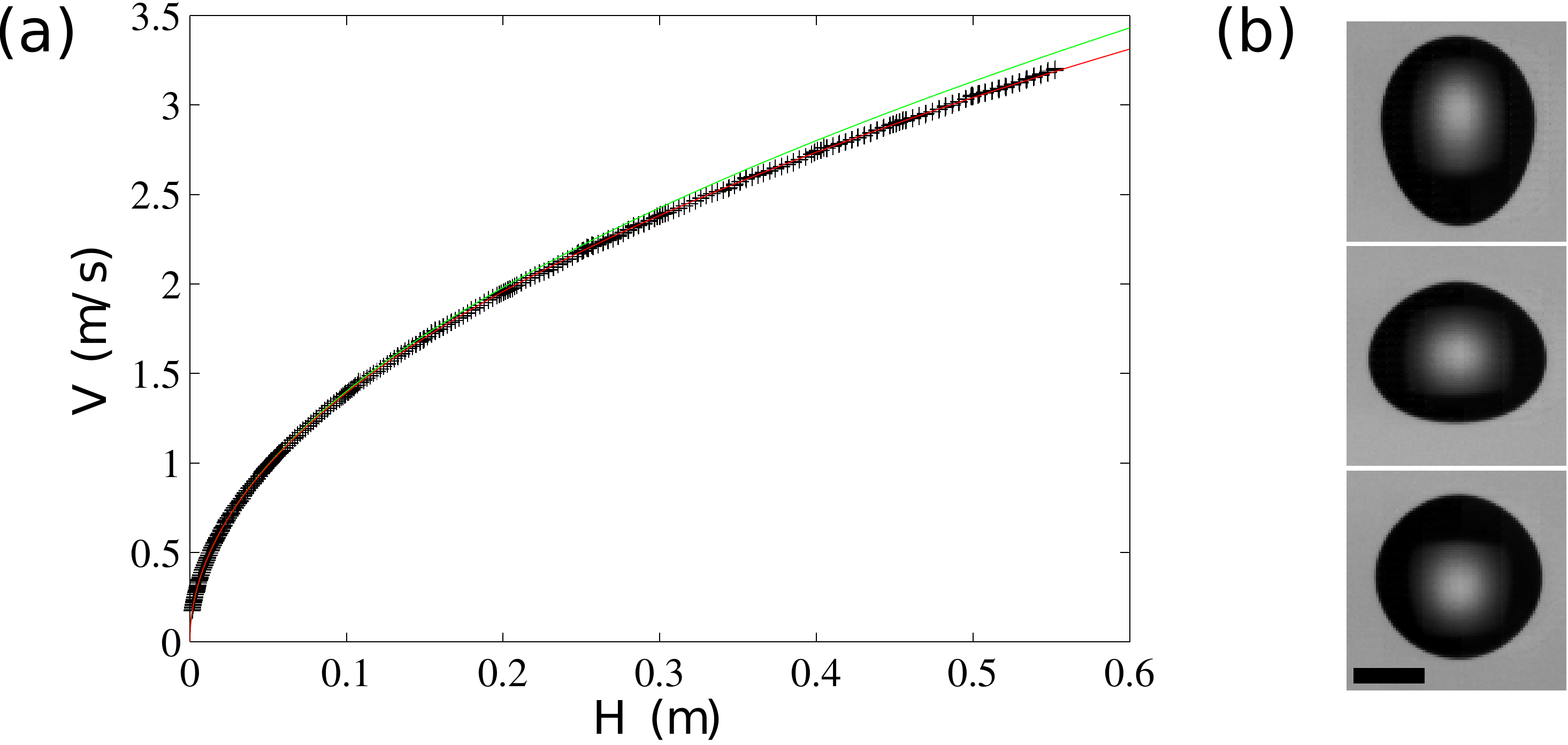}
\end{center}
\caption{(a) Drop velocity $V$ vs. falling height $H$.
The green line corresponds to $\sqrt{2gH}$, while the red line is our fitting equation.
(b) Typical drop shapes in air: prolate ($\alpha > 1$, $H = 31.2~cm$, $Re = 11\,300$),
oblate ($\alpha < 1$, $H = 41.8~cm$, $Re = 12\,900$)
and close to spherical ($\alpha \simeq 1$, $H = 44.5~cm$, $Re = 13\,300$).
The scale bar is 2 mm long.
}
\label{Fig_011}
\end{figure}

The drop velocity $V$ was characterized in a separate set of experiments.
It was then modeled by the velocity of a sphere experiencing constant drag:
\begin{equation}
V = V_T  \sqrt{1 - \exp{\left(-2g\left(h-D-h_0\right)/V_T^2\right)}},
\end{equation}
where gravity is $g = 9.81$ m/s$^2$, with the fitting parameters $V_T = 9.11$~m/s and $h_0 = 2.1$~mm.
$V_T$ corresponds to the terminal velocity of the drop,
and $h_0$ to the pinch-off length of the drop when it separates from the nozzle.
Here $h$ is the measured distance from the nozzle tip to the undisturbed pool surface,
whereas the adjusted height is defined as $H = h - D - h_0$.
Figure~\ref{Fig_011}(a) shows that the measured values of $V$ are less than 0.8\% away from the formula, 
for our impact heights 2.5~cm $<$ $H$ $<$ 55~cm.
This estimate of $V_T$ is slightly higher than the experimental observations of \citet{Gunn1949},
which could be due to the drop oscillations before reaching a final oblate shape.
We can calculate the falling time of the drop from the falling height as:
$t = (V_T/g) \argch \left[ \exp \left( g H/V_T^2 \right) \right]$.

\begin{table*}
\begin{center}
  \begin{tabular*}{0.85\textwidth}
     {@{\extracolsep{\fill}}|c|ccccc|}
\hline
Liquid & $\rho$ [g/cm$^3$] & $\mu$ [cP] & $\nu$ [cSt] & $\sigma$ [dyne/cm] & $n$ \\
\hline
Distilled water & 0.996 & 1.004 & 1.008 & 72.1 & 1.333 \\
Ethanol         & 0.789 & 1.19  & 1.51  & 23.2 & 1.363 \\
Methanol        & 0.793 & 0.593 & 0.748 & 22.5 & 1.339 \\
\hline
\end{tabular*}
\caption{Properties of the different liquids used in the pool.
Here $\rho$ is the liquid density; $\mu$ is the dynamics viscosity;
$\nu$ is the kinematic viscosity, $n$ the refractive index at $\lambda$~=~532~nm,
and $\sigma$ the surface tension.
The drop is always water.
}
\label{table01}
\end{center}
\end{table*}

\begin{figure}
\begin{center}
\includegraphics[width=0.88\linewidth]{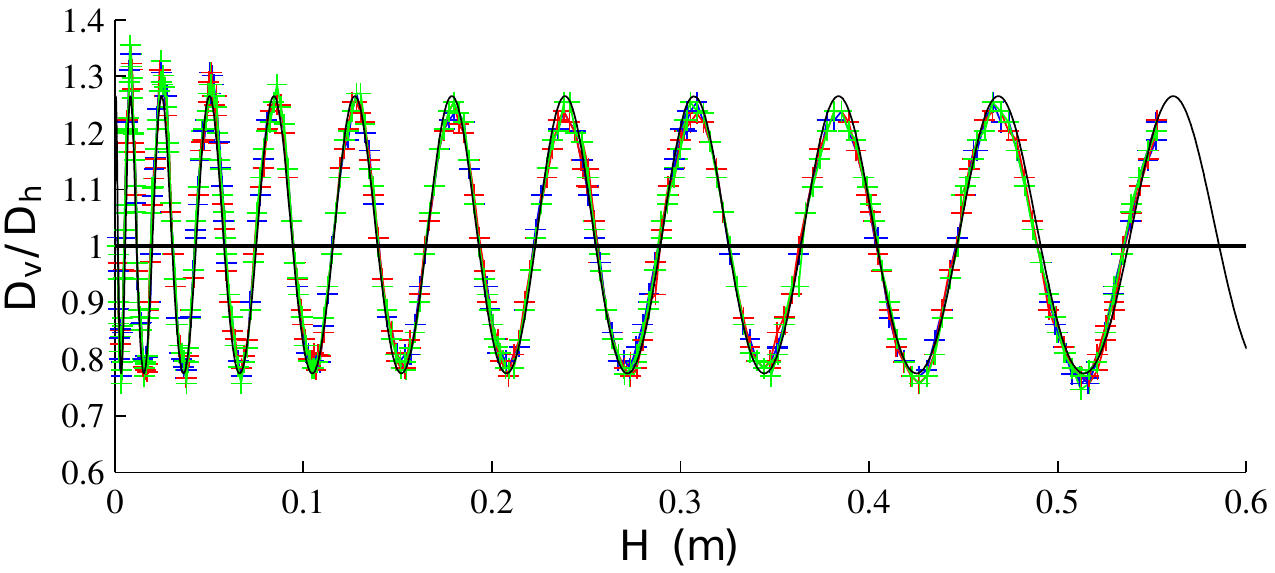}\vspace{0.1 in}\\
\end{center}
\caption{Drop aspect ratio $\alpha = D_v/D_h$ vs. $H$.
The black oscillating line corresponds to the fitted equation \eqref{eq:AvsT}.
}
\label{Fig_012}
\end{figure}

As $D$ is larger than the capillary length for water, $l_c = \sqrt{\sigma / (\rho g)}=$ 2.7 mm, 
the water drop shows large oscillations that can affect the details of the impact dynamics (see Fig.~\ref{Fig_011}(b)).
The axisymmetric vertical oscillations of the drop can be estimated by the dominant mode (\citet{Rayleigh1879}, \citet[][$\mathsection$ 275]{Lamb1975}), giving a radius:
\begin{equation}
R(t, \theta) = R_0 \left[ 1 + a \cos \left( \omega t + \phi \right) P_2 \left( \cos \theta \right) \right],
\end{equation}
where $P_2 \left( x \right) = \left( 3 x^2 - 1 \right) / 2$ is the Legendre polynomial of degree 2,
and $\theta$ is the polar angle in the spherical coordinate system.
The aspect ratio between the vertical and horizontal diameters of the drop can thus be written:
\begin{equation}\label{eq:AvsT}
\alpha = \frac{D_v}{D_h} = \frac{1 + a \cos \left( \omega t + \phi \right)}{1 - \frac{a}{2} \cos \left( \omega t + \phi \right)}
\end{equation}
We determine $a$, $\omega$ and $\phi$ as fitting parameters:
$a = 0.162$, $f = \omega/(2\pi) = 33.2$~Hz and $\phi = -132\degree$.
The oscillation frequency is only 2\% lower than the inviscid theoretical value 
$f_D = (4/\pi) \sqrt{\sigma/(\rho D^3)} = 33.9$~Hz.
The typical time of bubble-ring entrapment, 50 $\mu$s, is only 0.17\% of the oscillation period.
%100 $\mu$s, is only 0.33 \% of the oscillation period.
Therefore the drop shape can be considered frozen during the entrapment.
This fitting is then used to get the aspect ratio from the falling height in the experiments.
We have neglected viscous damping of the dominant mode in this estimate of drop oscillations.
The characteristic time of this damping can be estimated as 
$\tau = D^2 / (20 \nu) = 1.08$~s \citep[][$\mathsection$ 355]{Lamb1975},
In the overall falling time studied here ($\simeq 0.35$~s),
viscous effects can be estimated to reduce the amplitude of the dominant mode by 27\%.
It is therefore too short to damp the oscillations significantly, as is observed in Fig.~\ref{Fig_012}.

\subsection{Numerical method}
\label{sec:Numerics}

We use the open-source software Gerris (http://gfs.sf.net; \citet{Popinet2003, Popinet2009, Agbaglah2011}),
using the \textit{Volume-Of-Fluid} method, to perform axisymmetric simulations of the drop impacts.
The liquid from the drop and the pool are identified with different markers
(drop: red, pool: blue, air: light green).
The adaptive mesh is refined dynamically based on the distance to the interface,
vorticity magnitude and geometric conditions.
The interface is refined uniformly at the maximum level in the simulations.
The bubbles and droplets with area less than 10 cells are removed during the computation,
as their dynamics cannot be captured accurately.
It represents an effective cutoff diameter of $D_{cut}$ = 3.57 cells.

The simulations are started with the drop $0.1\, R$ above the pool,
where $R = D/2$ is the drop radius.
Non-dimensional time is defined as $t^* = t / \tau$, where $\tau = D/V$.
The origin of time is taken when the undisturbed sphere would first contact the pool.
The drop is kept at a constant effective diameter $D = 4.6$ mm.
Air has a viscosity of $\mu_{a} = 1.81 \times 10^{-2}$ cP and density $\rho_{a} = 1.21$~kg/m$^3$.
The liquid is water for both the drop and the pool, with viscosity $\mu = 1$ cP,
density $\rho = 1000$~kg/m$^3$ and surface tension $\sigma = 72$~mN/m.
Gravity is included as $g = 9.81$ m/s$^2$.
We do not take into account the different properties of the bottom liquid in the simulation,
and therefore do not include any Marangoni or variable-density effects between the two liquids.
More details about the adaptive grid refinement can be found in \citet{Popinet2003} and \citet{Thoraval2012}.

\begin{figure}
\begin{center}
\includegraphics[width=\linewidth]{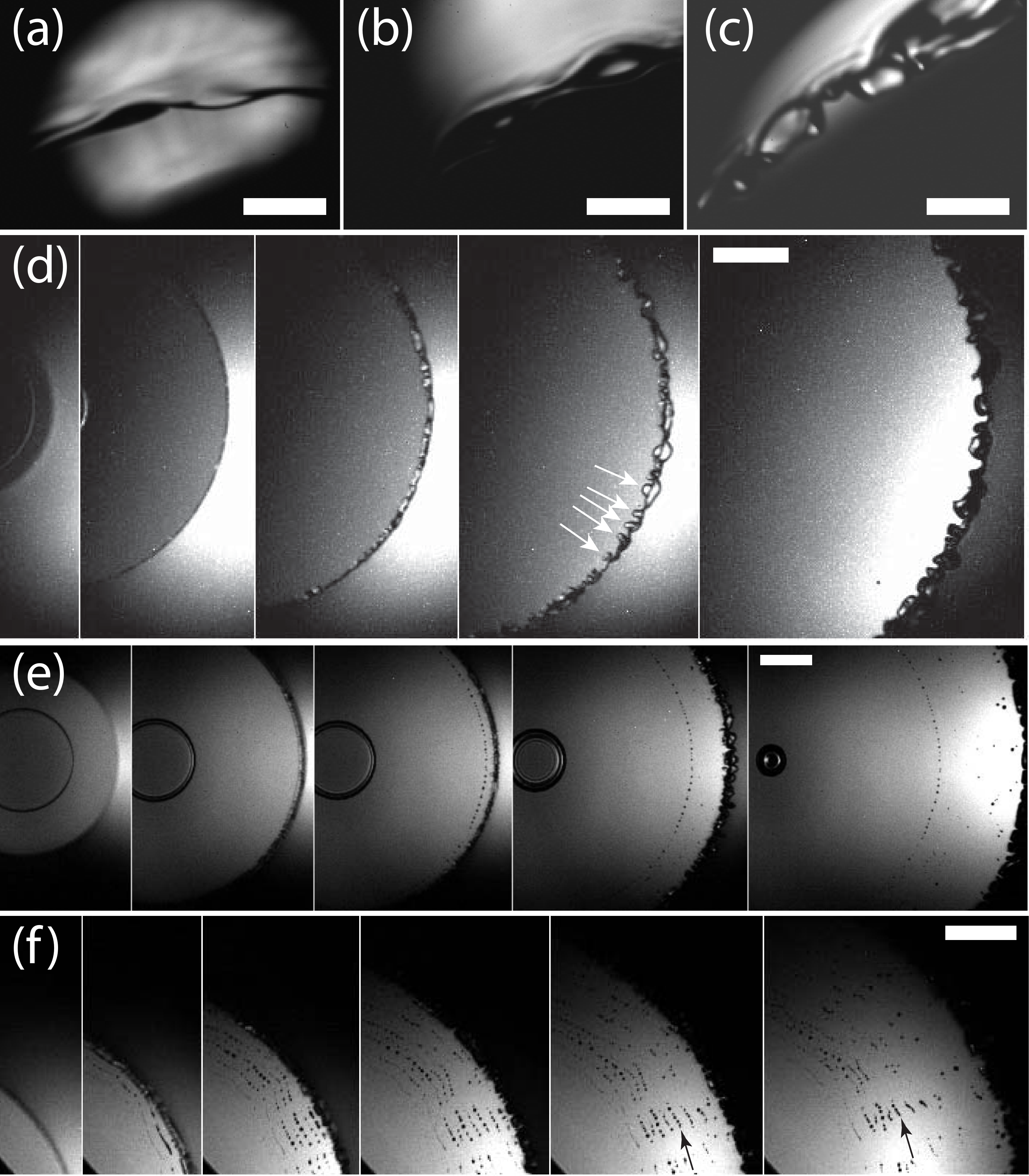}\vspace{0.1 in}
\end{center}
\caption{Early contact of a water drop impacting a $\delta = 250\; \mu$m deep layer of water.
(a-c) No bubble entrapment is observed at low impact velocities.
An azimuthal instability develops on the ejecta,
with a wavelength decreasing with increasing Re.
(a) $Re = 3\,610$, $We = 39$, $K = 2\,360$, $\alpha$ = 0.94,
(b) $Re = 4\,400$, $We = 58$, $K = 3\,860$, $\alpha$ = 1.17,
(c) $Re = 8\,980$, $We = 90$, $K = 6\,640$, $\alpha$ = 0.86.
(d) For an intermediate impact velocity,
individual micro-bubbles can be entrapped.
Frames are shown at 1, 13, 18, 25 \& 46 $\mu$s after the first contact.
$Re = 11\,400$, $We = 394$, $K = 42\,100$, $\alpha$ = 1.05.
(e) For slightly higher impact velocity,
the drop entraps one bubble ring and isolated bubbles,
shown at $t=$ 3, 11, 13, 21 \& 40 $\mu$s.
$Re = 13\,300$, $We = 535$, $K = 61\,800$, $\alpha$ = 0.98.
(f) Multiple bubble rings. 
Frames are shown at about 5, 9, 13, 17, 21 and 32 $\mu$s after first contact.
$Re = 12\,900$, $We = 506$, $K = 57\,600$, $\alpha$ = 0.80.
The scale bars are all 200 $\mu$m long.
See also supplemental videos.
}
\label{Fig_02}
\end{figure}

\begin{figure}
\begin{center}
\includegraphics[width=\linewidth]{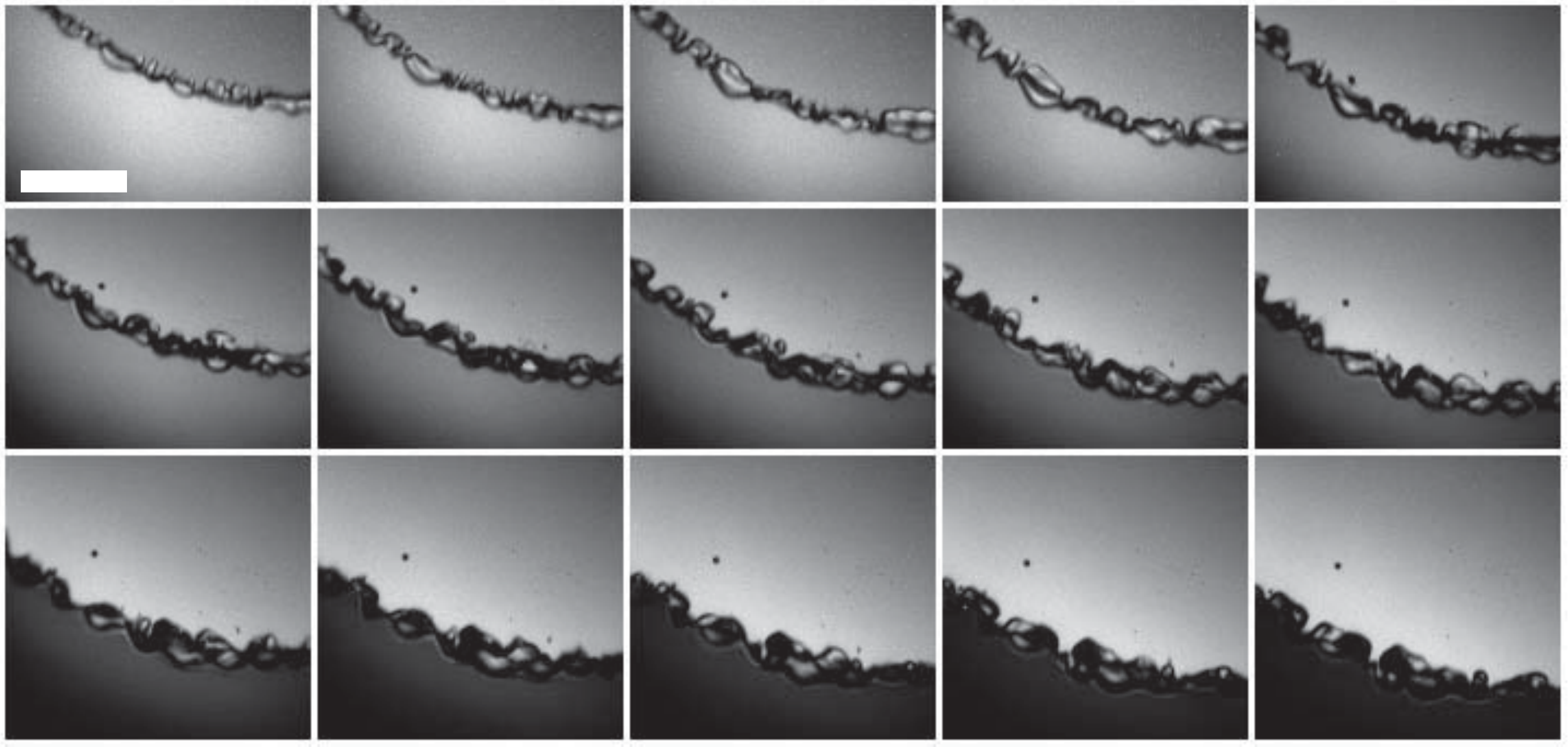}
\end{center}
\caption{Details of individual micro-bubbles entrapments
in the same conditions as in Fig.~\ref{Fig_02}(d).
Frames are shown 2~$\mu$s apart.
The top row shows the dynamics leading to the entrapment
of a $\simeq 10$~$\mu$m diameter bubble.
On the second row, two smaller bubbles of diameter $\simeq 4$ and $6$~$\mu$m 
are separating from the right part of the edge.
While the first one stays behind the edge, and can be seen in the last frame,
the second one is re-absorbed into the neck in the last row.
The scale bar is 200~$\mu$m long.
}
\label{Fig_022}
\end{figure}

\section{Results and discussion}
\label{sec:ResultsAndDiscussions}

\subsection{Isolated bubbles and multiple bubble rings for a water pool}
\label{sec:WaterRings}
We start by looking at the impact of the water drop onto a water layer.
Figure~\ref{Fig_02} shows the early evolution of the outer neck contact of the drop with the pool.
The contracting inner air disk is visible on the left side of the images in panels (d) and (e).
Note that we are only looking at the early contact when the neck has not reached the size
of the drop, as shown in the sketch in Fig.~\ref{Fig_01}(b).
The radius of the neck in the last panel of Fig.~\ref{Fig_02}(d) has only reached 37\% of the drop radius.

Figure~\ref{Fig_02} shows that
even for low impact velocities the neck region between the drop and the pool does not remain
smooth and axisymmetric, but develops azimuthal undulations.
For the lowest impact velocities these undulations have long wavelengths and do not entrap bubbles,
see Fig.~\ref{Fig_02}(a-c).
However, with increased impact velocity $V$ the wavelength reduces and their amplitude grows more rapidly.
In Fig.~\ref{Fig_02}(d) these undulations appear first in the second panel and grow in amplitude during the radial motion,
but individual bumps saturate and are often being pulled back by surface tension.
The shapes are irregular, but we can glean a characteristic wavelength from the third panel in Fig.~\ref{Fig_02}(d),
giving $\lambda \sim 53 \; \mu$m, corresponding to 73 undulations around the periphery.
These undulations appear when the ejecta emerges, pulling local sheets of air under the ejecta
on both or alternating sides of it.
These local sheets can be pulled along with the ejecta base,
with only occasional bubbles entrained,
when these small azimuthal air discs make contact across the thin air layer,
as is shown in a longer sequence of frames in Fig.~\ref{Fig_022}.
Individual bubble entrapments can also occur in the troughs between the undulations.

\begin{figure}
\begin{center}
\includegraphics[width=0.70\linewidth]{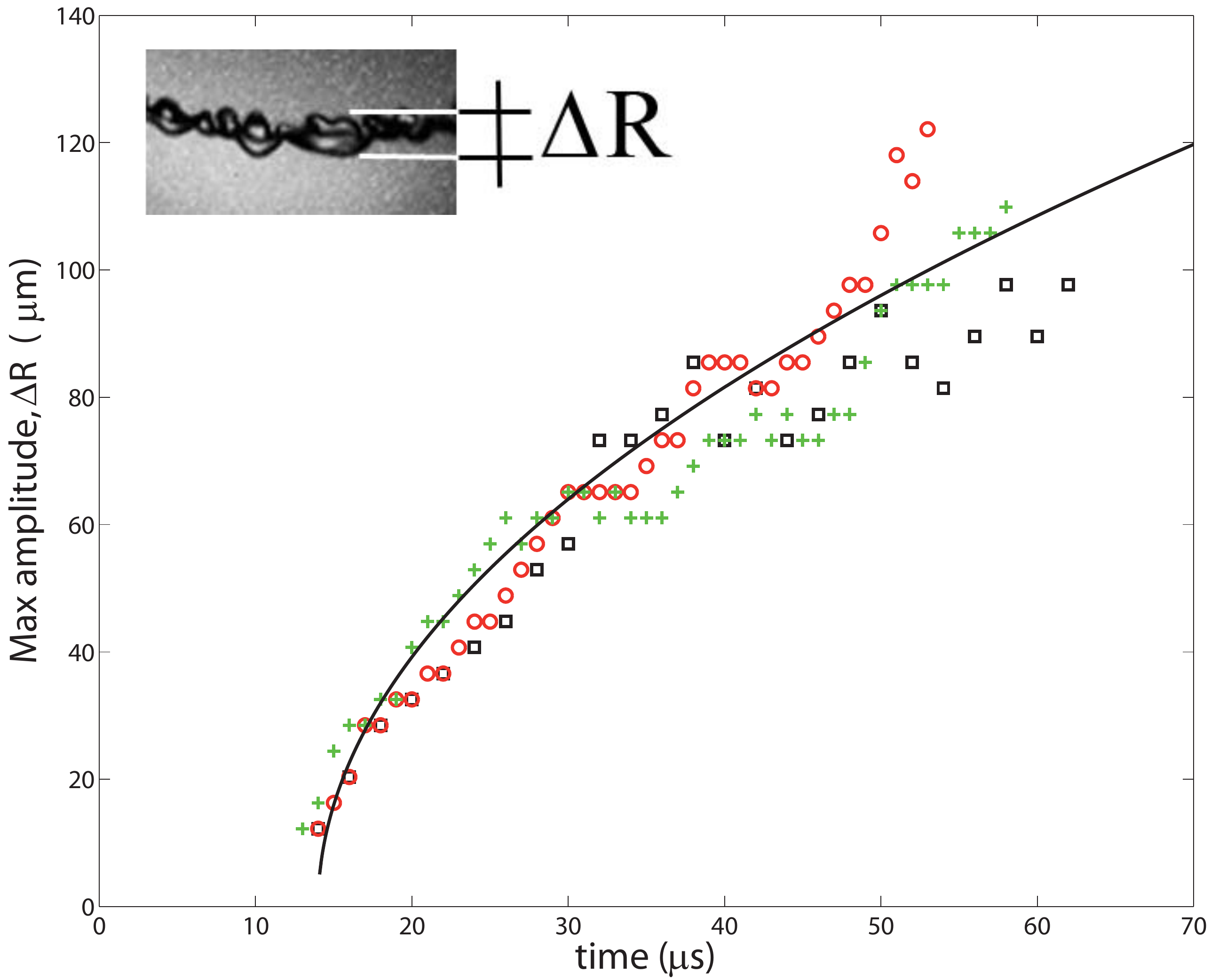}
\end{center}
\caption{The maximum amplitude of the undulations of the front in the neck region, for water drop onto water pool.
For the same conditions as Figs.~\ref{Fig_02}(d) and ~\ref{Fig_022}. Data from three realizations.
The solid line shows formula \ref{eq_L}, with $C=14$.
}
\label{Fig_023}
\end{figure}

In Fig.~\ref{Fig_023} we estimate the growth of the maximum undulation amplitude, measured between the troughs and peaks.
The growth slows down with radial distance.
For reference we plot a viscous length-scale
\begin{equation}
L_{\nu} = C \; \sqrt{\nu (t-t_o)}
\label{eq_L}
\end{equation}
suggesting inertia and viscous forces both play a role.
Here $t_o$ is the time of first observed undulations on the front.

Figure~\ref{Fig_022} shows that the characteristic azimuthal size of the undulations also grows during the radial
motion of the front, but this is more difficult to quantify.

It is curious that some air entrapment in breaking gravity waves has superficially similar appearance
(\cite{KigerDuncan2012}, their Fig. 11), but is clearly driven by a different mechanism and is 
three orders of magnitude larger in size.

For slightly higher impact velocity,
the entrapment of rather irregular bubble arcs begins.
Figure~\ref{Fig_02}(f) shows up to 10 such partial rings.
The average radial spacing of the adjacent bubble rings is $\simeq 26 \; \mu$m.
The air cylinders then break up into a row of bubbles through
surface-tension driven Rayleigh instability.
Bubbles are often shifted sideways in the azimuthal direction
during the radial spreading (arrows in Fig.~\ref{Fig_02}(f) and supplementary videos).
For a stationary hollow cylinder of diameter $d_b$ in an inviscid liquid,
the most unstable wavelength is $\lambda_m = \pi d_b / 0.484$.
The characteristic time scale of the exponential growth $\sim exp(t/\tau_{\sigma})$
of the breakup is given by $\tau_{\sigma} = 1.22 \sqrt{\rho r_b^3 / \sigma}$ \citep{Chandrasekhar1961}.
The radii of the bubble arcs for a water layer, in Fig.~\ref{Fig_02}(f), are $\sim $ 3 $\mu$m
and they break up in about $\sim$ 3 $\mu$s, which is $4\tau_{\sigma}$.

Therefore, the first bubble-rings entrapment for water occurs
around $Re \simeq 12\,000$ and $K \simeq 50,000$.
These values are consistent with numerical results of \citet{Thoraval2012},
where no bubble ring entrapment is observed for $Re = 10\,000$ and $K = 30\,000$,
and a row of bubble rings observed for $Re = 14\,500$ and $K = 74\,400$.
In the former case, the ejecta sheet is thicker, because of the stronger surface tension
effects on the ejecta sheet owing to the lower value of the splashing parameter $K$.
It is re-absorbed on the drop or the pool during the oscillations,
and no bubble ring entrapment is predicted.
However, in the latter case, at higher $K$, the ejecta sheet is thinner,
and the oscillations entrap a row of bubble rings at the core of vortex rings
when it impacts onto and connects with the drop or the pool.

However, the comparison of Fig.~\ref{Fig_02}(e) and (f) shows that more bubble rings can be observed
at a slightly lower $Re$ and $K$.
This suggests that the $Re$ number of the impact is not enough to characterize the bubble-rings entrapment.
We will show in \S\ref{sec:PoolDepthDropShape} the critical effect of the drop shape.
Moreover, the azimuthal instabilities also affect the air entrapment,
and individual bubble entrapments have been observed at slightly lower $Re$ in
Figs.~\ref{Fig_02}(d) \& \ref{Fig_022}.

Note that in the work of \citet{CastrejonPita2012}, the vortex street is
observed for conditions similar to Fig.~4(f) of \citet{Thoraval2012}, where no bubble-ring entrapment was predicted.
Considering the large diameter of the drop they are using, it is not clear if the
bubbles they observe are part of a bubble ring or isolated bubbles.
They are also looking at a larger view and longer time evolution,
that could be out of the field of view used in the current investigation (see perspective in Fig.~\ref{Fig_01}).
However, their alternating vortices are a new observation,
clearly different from the isolated vortex rings produced by
much lower impact velocities, see \cite{PeckSigurdson1994}.

\subsection{Bubble rings for miscible liquids}
\label{sec:MiscibleRings}

\begin{figure}
\begin{center}
\includegraphics[width=\linewidth]{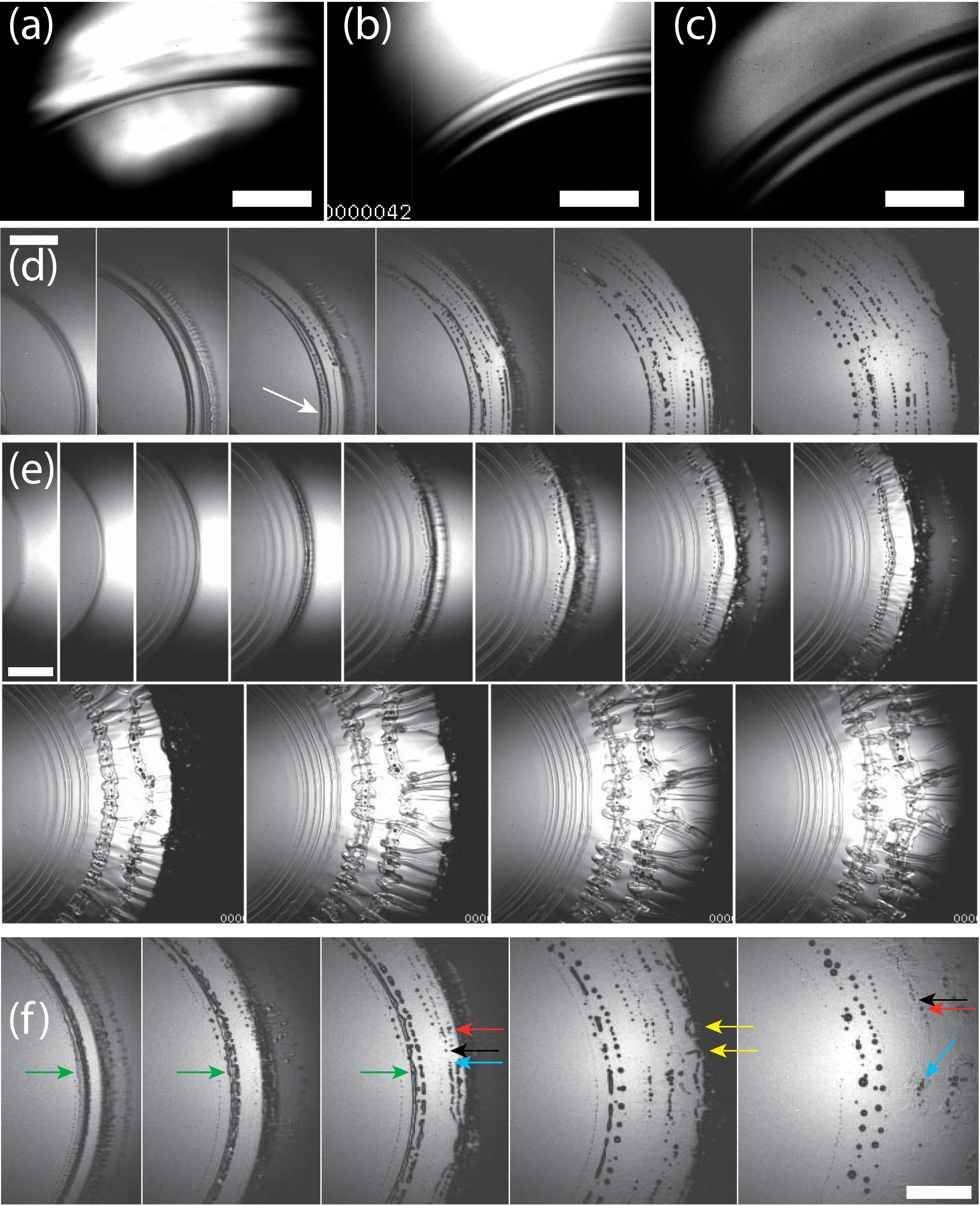}
\end{center}
\caption{Early contact of a water drop impacting a pool of a miscible liquid (ethanol or methanol).
(a-c) No bubble entrapment is observed at low impact velocities on a thin film of ethanol
($\delta \simeq$ 250~$\mu$m).
(a) $Re = 3\,610$, $\alpha$ = 0.94,
(b) $Re = 4\,400$, $\alpha$ = 1.17,
(c) $Re = 8\,980$, $\alpha$ = 0.86.
(d) Bubble rings for a water drop impacting onto a methanol layer
($Re=12\,900$, $\alpha = 0.80$, $\delta = 50\; \mu$m).
Frames are shown at 5, 9, 11, 15, 20 \& 33 $\mu$s after first contact.
(e) First oscillations of the ejecta sheet, followed by entrapment of bubble rings,
for a film of ethanol ($Re=13\,300$, $\alpha = 0.98$, $\delta = 250\; \mu$m).
Azimuthal instabilities appear in the ejecta sheet, before its rim detaches in a liquid toroid.
The first 8 frames are shown 4~$\mu$s apart, and then 20~$\mu$s.
(f) Bubble entrapment dynamics for impact on a methanol film
($Re=12\,900$, $\alpha = 0.80$, $\delta = 50\; \mu$m).
The first frames shows the entrapment of a superposition of air sheets,
later breaking in patches and then in micro-bubbles (green arrows).
Three bubbles are identified in the third frame by red, black and blue arrows.
Bubble arcs with legs in the radial direction are identified by the yellow arrows.
Frames are shown 3, 6, 15 and 60~$\mu$s after the first one.
The scale bars are all 200~$\mu$m long.
}
\label{Fig_04}
\end{figure}

\begin{figure}
\begin{center}
\includegraphics[width=1.0\linewidth]{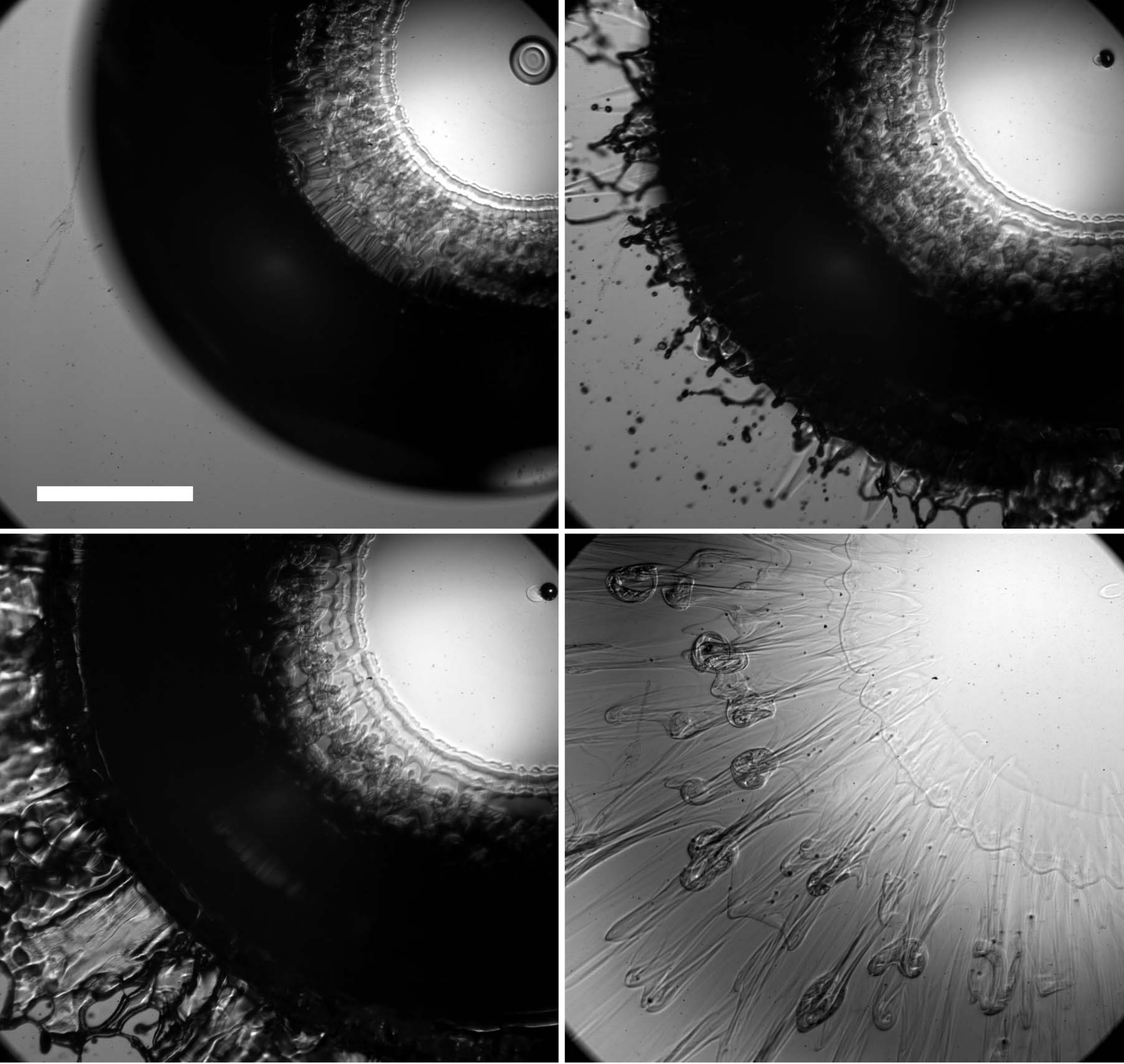}
\end{center}
\caption{Larger view of the drop impact onto ethanol, 
at times $t\sim$ 50, 150, 250 \& 1150 $\mu$s after first contact 
($Re= 14\,500$, $\alpha = 0.94$, $\delta \simeq 125 \; \mu$m).
Frames taken from videos using Photron SA-5 at 10,000 fps, with a 1~$\mu$s exposure time.
The central bubble has drifted out of the image in the last frame.
The scale bar is 1 mm.
}
\label{Fig_10}
\end{figure}

The bubble ring entrapment becomes more regular for the impacts onto ethanol and methanol pools (Fig.~\ref{Fig_04}),
perhaps due to the lower surface tension of these liquids (Table~\ref{table01}).
Contrary to the impacts onto water films, no azimuthal instability develop on the neck of the ejecta,
which remains perfectly smooth in Fig.~\ref{Fig_04}(a-c).
At higher impact velocities bubble rings are entrapped.
Figure~\ref{Fig_04}(d) shows at least 10 bubble-rings.
Many of them are entrapped axisymmetrically over the entire image view, 
which can span around 90$^o$ angular sector.

In some instances thin ribbons of air are entrapped and subsequently breakup into sub-rings and thereafter
bubble rings, as highlighted by arrow in Fig.~\ref{Fig_04}(d).
The frames in Fig.~\ref{Fig_04}(f) detail this sequence of air entrapments (green arrows).
The air sheet first breaks into smaller patches, later contracting into bubbles.
In the third frame, some bubbles are observed at the same radial location as another air cylinder.
This supports the mechanism that air can be entrapped both above and below the ejecta sheet,
as was shown in the numerical simulations of \cite{Thoraval2012}.

In a similar way as for water pools,
Fig.~\ref{Fig_04}(f) also demonstrates strong sideways motions of bubbles.
Three bubbles are identified in the third frame by red, black and blue arrows.
Their corresponding location is marked by the same colored arrows in the last frame.
Comparison between the black and red arrows shows that this sideways motion
can be of different strength for adjacent rings.
This bubbles motion in the azimuthal directions results in their clustering at isolated locations.
Bubble arcs with legs extending in the radial direction towards the neck, are identified by the yellow arrows.

Figure~\ref{Fig_10} shows a wider view of an impact taken at a lower frame rate of 10,000 fps but larger pixel
area of $896 \times 848$ px, using the Photron SA-5 CMOS camera.  Each frame is frozen with a 1 $\mu$s exposure.
The 100 $\mu$s interframe time only shows us snapshots of the phenomenon,
putting the earlier figures in perspective, with most of the earlier sequences occurring before the first image.
In the second frame multitude of splashed droplets appear from underneath the shadow of the drop,
with some larger droplets planing on the pool surface, leaving behind narrow capillary wedges.
The first panel shows a smooth central regions, followed by a convoluted interface,
suggesting the stirring by the three-dimensional vortical structures (see \S\ref{sec:Tangle}).  
Similar stirring can be inferred from the side shadowgraph imaging in \cite{CastrejonPita2012} (their Fig. 4).
The final panel in Fig.~\ref{Fig_10} shows numerous isolated bubbles which have been redistributed by the vortical motions.
The bubbles are mostly concentrated within the mushroom-like remnants of the vortical structures.

\begin{figure}
\begin{center}
\includegraphics[width=1.0\linewidth]{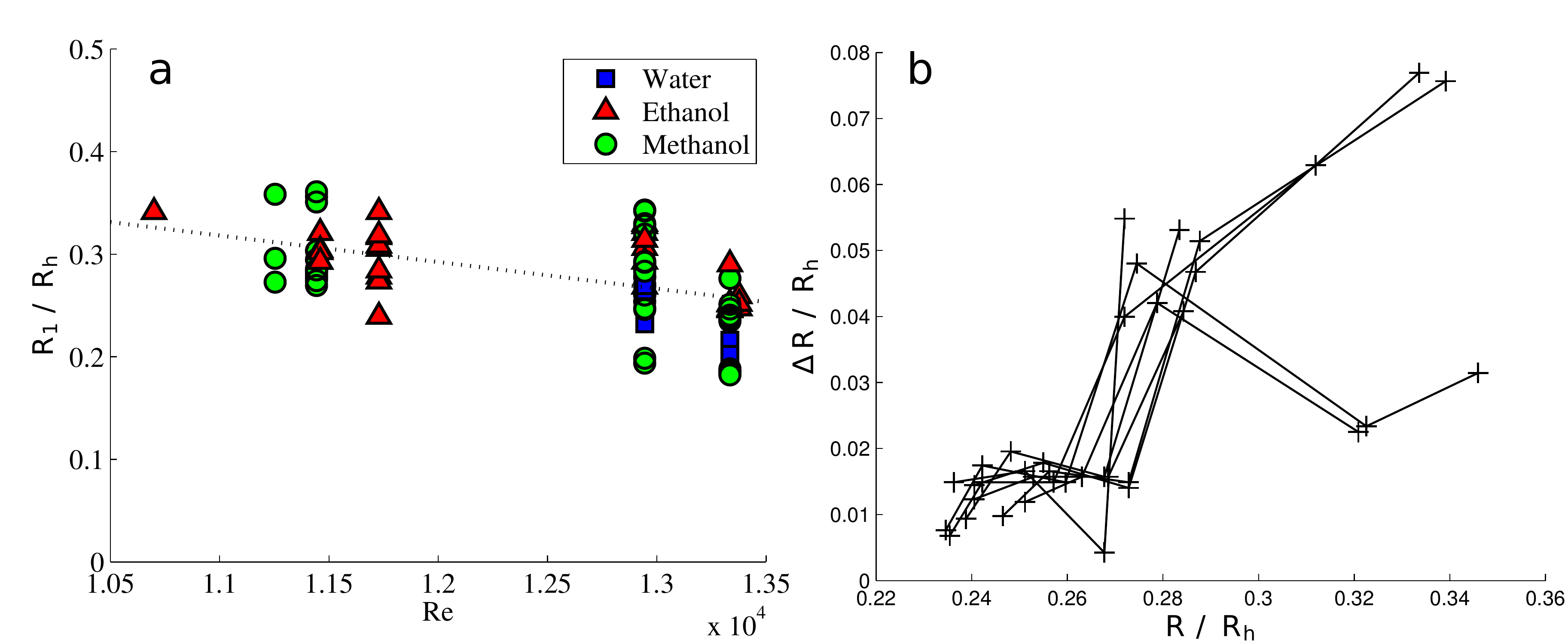}
\end{center}
\caption{(a) First onset of bubble rings for water drop impacting onto a
water, ethanol or methanol layer of different depths (see also Fig.~\ref{Fig_07_2}(b)).
(b) Spacing of adjacent rings $\Delta R$, vs radial distance $R$ from the impact center.
Data from 8 different realizations for a methanol pool ($Re=13\,300$, $\alpha = 0.98$, $\delta = 250\; \mu$m).}
\label{Fig_05}
\end{figure}

\subsection{First onset, number and spacing of bubble rings}
\label{sec:Radius}
The first contact entraps a central air disk and forms a rapidly expanding outer liquid edge.
Bubble rings are then formed as observed above.
Figure~\ref{Fig_05}(a) looks at the radial location where the first bubble ring is entrapped.
We normalize the radius of the first ring $R_1$ 
with the horizontal drop radius when it first contacts the pool, $R_h = D_h/2$.
The data shows large spread, but an overall trend is for the onset to occur earlier for
larger impact $Re$. The lowest entrapment radius is 0.18, similar to the 0.2 limit
observed by \citet{Thoroddsen2002} for the onset of the ejecta sheet.
This onset radius of the ejecta is also in agreement with the inviscid numerics of \cite{WeissYarin1999}.

Figure~\ref{Fig_05}(b) shows the distance between the adjacent bubble rings 
measured for numerous identical impact conditions.
The spacing of the rings tends to increase with distance and the entrapped bubbles become larger. 

\begin{figure}
\begin{center}
\includegraphics[width=0.98\linewidth]{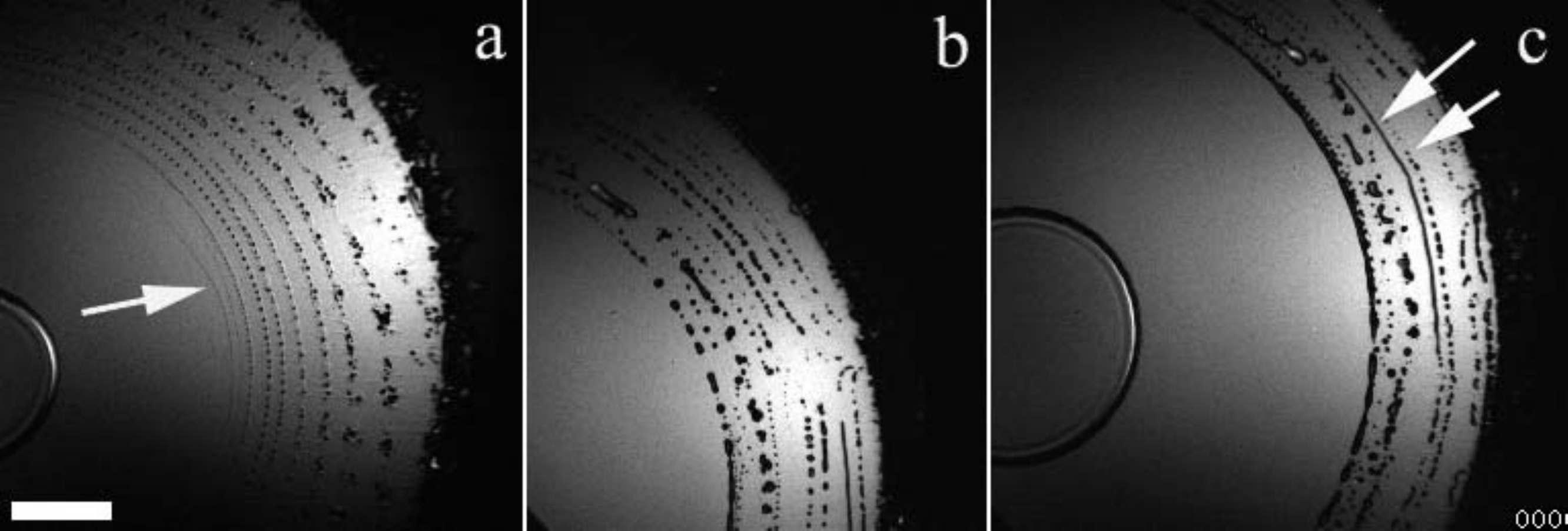}\vspace{0.05in}
\includegraphics[width=0.98\linewidth]{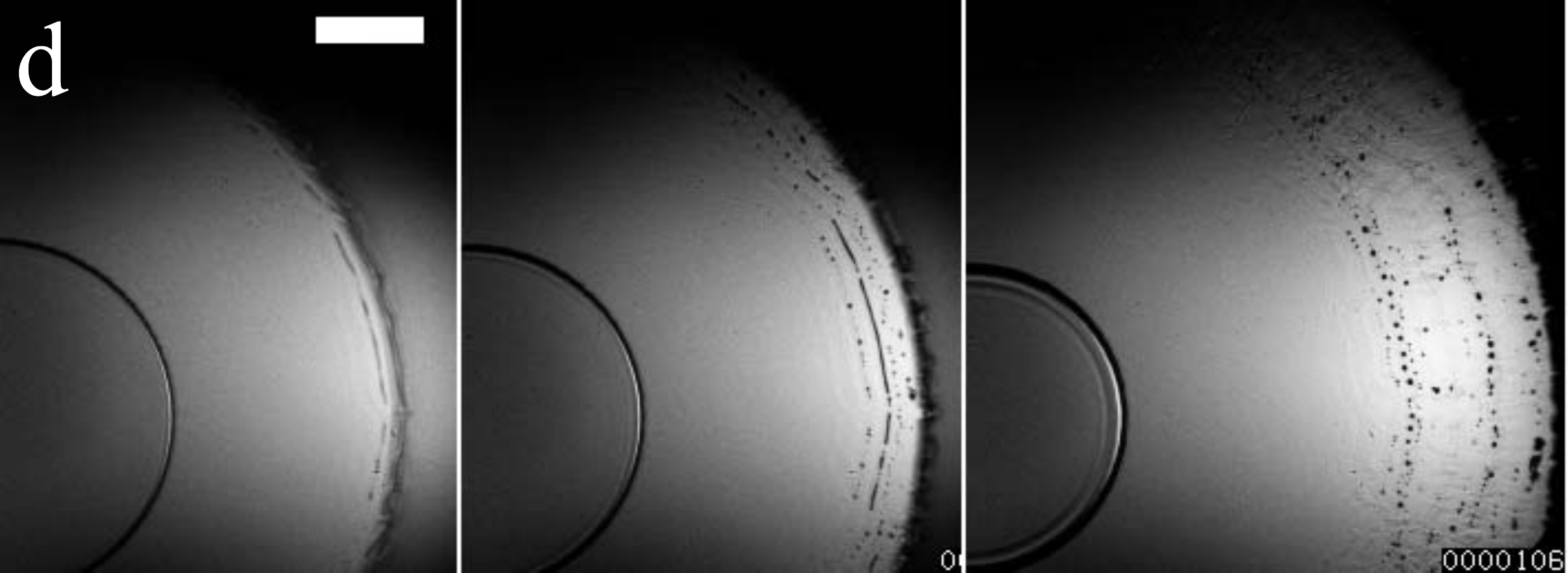}\vspace{0.05in}
\end{center}
\caption{Effect of pool depth on the bubble rings
for water drop impacting onto methanol,
with $\delta \simeq$ 25 (a), 50 (b) and 75 $\mu$m (c)
($Re=12\,900$, $\alpha = 0.80$).
The frames are all shown 24 $\mu$s after the first contact in (a-c). 
(d) Deeper methanol pool with $\delta \simeq$ 500 $\mu$m.  Frames shown at 10, 14 \& 28 $\mu$s after first contact.
The scale bars are 200 $\mu$m long.  See also supplemental videos.}
\label{Fig_06}
\end{figure}

\subsection{Effect of pool depth and drop shape}
\label{sec:PoolDepthDropShape}

To ascertain the influence of the pool depth we systematically vary the layer thickness $\delta$,
from about 25 $\mu$m to 1~mm.
Figure~\ref{Fig_06}(a-c) compares the bubble rings for the 3 smallest pool depths $\delta$.
The ring structures are qualitatively similar in all cases, 
but the shallowest pool shows the earliest and finest bubble rings,
some of which are sub-pixel in diameter.
The second ring in Fig.~\ref{Fig_06}(a) allows us to measure the separation of micro-bubbles,
giving $\lambda = 8.8\; \mu$m, suggesting a diameter of the original air torus $d_{tor} \simeq \; 1.4 \; \mu$m.
The earliest ring appears even smaller, arrow in Fig.~\ref{Fig_06}(a).
Figure \ref{Fig_06}(d) shows fewer by qualitatively similar ring entrapment, for much thicker layer.

\begin{figure}
\begin{center}
\includegraphics[width=0.88\linewidth]{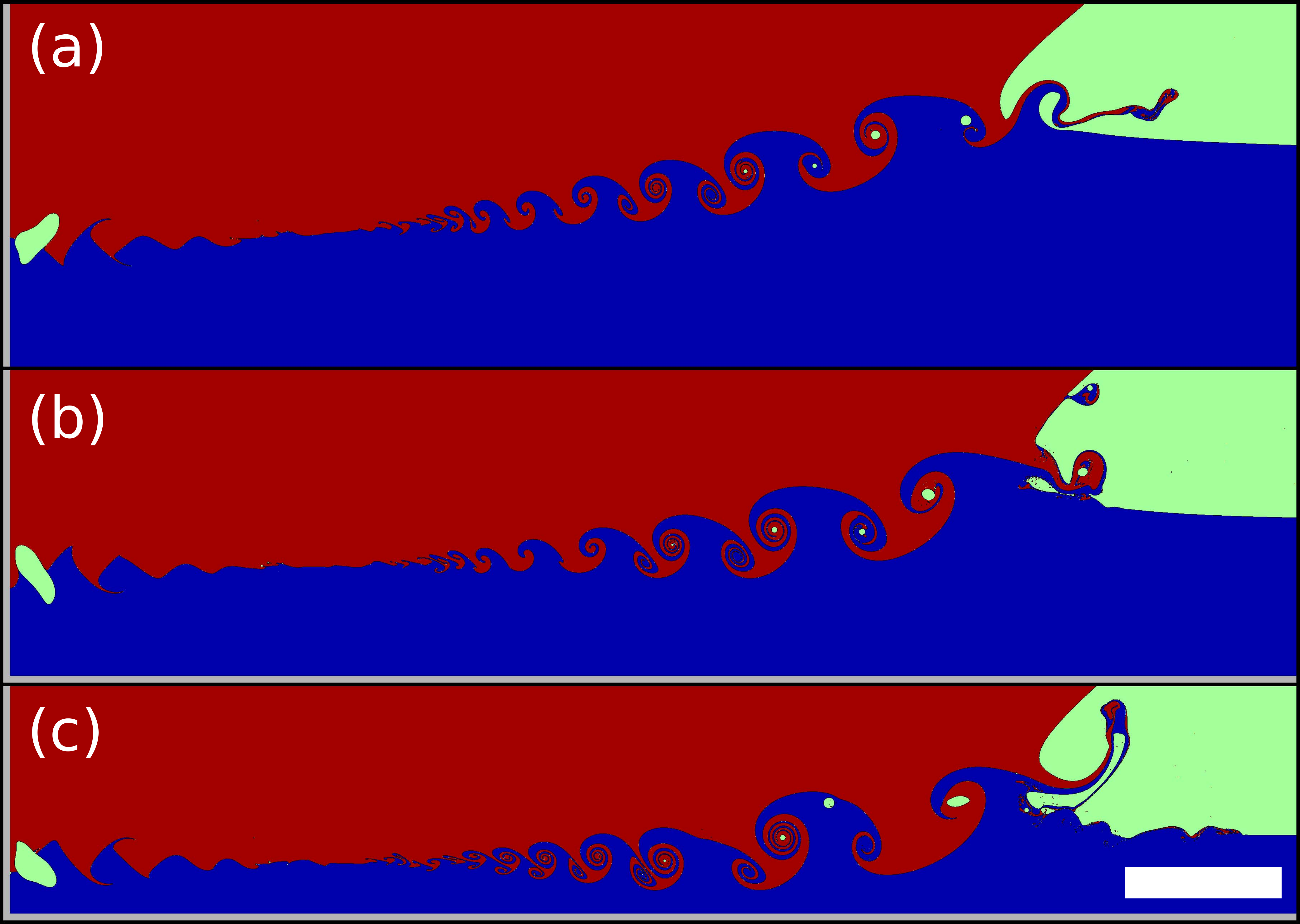}\vspace{0.1in}
\end{center}
\caption{Effect of pool depth on the vortex street at t$^*$ = 0.0526, 
for $Re = 13\,700$.
Pool depth $\delta$ = 800~$\mu$m (a), 200~$\mu$m (b), 100~$\mu$m (c).
The top of the 3 images are at the same location, relative to the original pool surface.
The pool depth is larger than shown in the image in (a),
but is completely included for (b) \& (c), where the bottom is indicated in gray.
The maximum and minimum level of refinement in the domain 
are respectively 12\,500 and 778 cells per drop diameter and $D_{cut}$ = 1.3~$\mu$m.
The bar is 200~$\mu$m long.
We can observe that one part of the ejecta sheet is climbing on the drop in (b),
while the main ejecta sheet continues to emerge below.
This is similar to what was observed numerically by \citet{Thoraval2012} and experimentally by \citet{Zhang2012}.
The formation of this higher part of the ejecta sheet can be observed in the Supplemental Video.
It then merges with the drop in this case.
}
\label{Fig_07}
\end{figure}

Figure~\ref{Fig_07} shows numerical results for three different pool depths,
where the two-liquid interface is highlighted by coloring the drop and pool differently.
It shows the drop penetrating further into the pool for larger $\delta$.
The vortex street is therefore constrained in a shallower region for shallower pools,
and therefore develops more horizontally.
This constraining effect increases the maximum liquid velocity by as much as 20\%,
as shown in Fig.~\ref{Fig_07_2}(a).
The first bubble ring is entrapped at $R_{1}/R$ of respectively 0.37, 0.31 and 0.28
for pool depths $\delta$ = 800, 200 and 100~$\mu$m respectively.
This confirms the previous experimental observation of Fig.~\ref{Fig_06}
that earlier rings are observed for shallower pools.
Figure \ref{Fig_07_2}(b) also shows this experimentally in a more systematic way,
but the difference is not very pronounced.
The radial location of this first entrapment is also consistent between experimental
and numerical observations, even though numerical simulations only considers one liquid.

\begin{figure}
\begin{center}
\includegraphics[width=\linewidth]{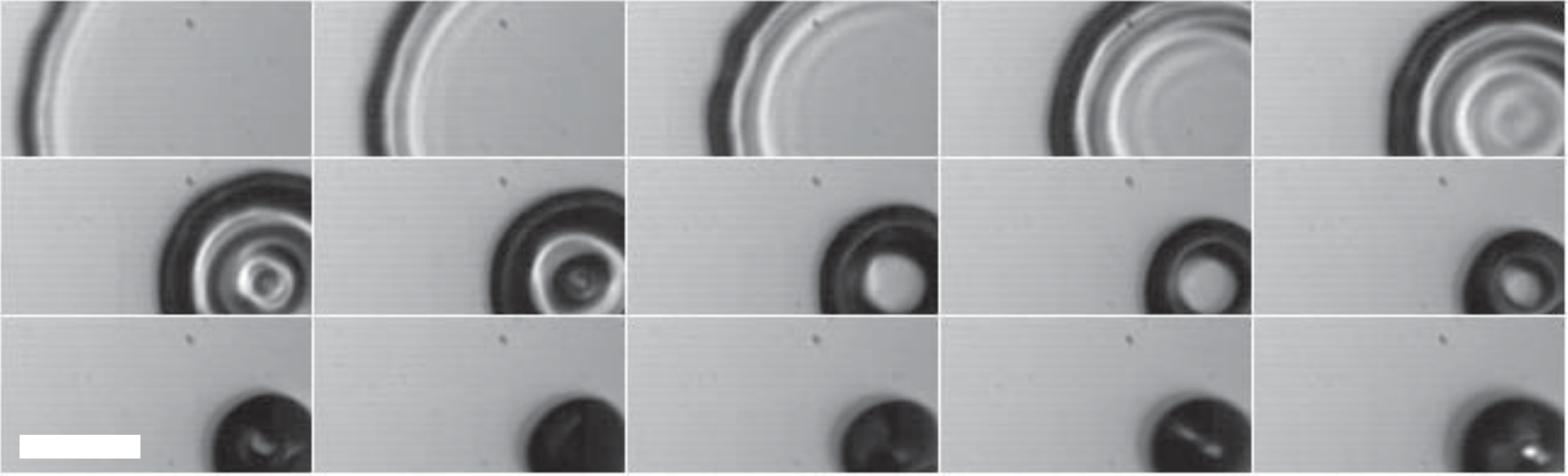}
\end{center}
\caption{Contraction of the central air disk for a water drop impacting on a thin water film
($Re=11\,600$, $\alpha = 0.86$, $\delta = 250\; \mu$m).
The convergence of capillary waves punctures the center of the disk,
producing an air toroid in the third frame of the second row, later contracting into a spherical bubble.
Similar formation of a central air toroid can be observed in the supplemental video of
Fig.~\ref{Fig_02}(e), as well as in other water drop impacts on ethanol or methanol films.
Frames are shown 7.1~$\mu$s apart, with an exposure time of 1~$\mu$s.
}
\label{Fig_08_2}
\end{figure}

\begin{figure}
\begin{center}
\includegraphics[width=1.0\linewidth]{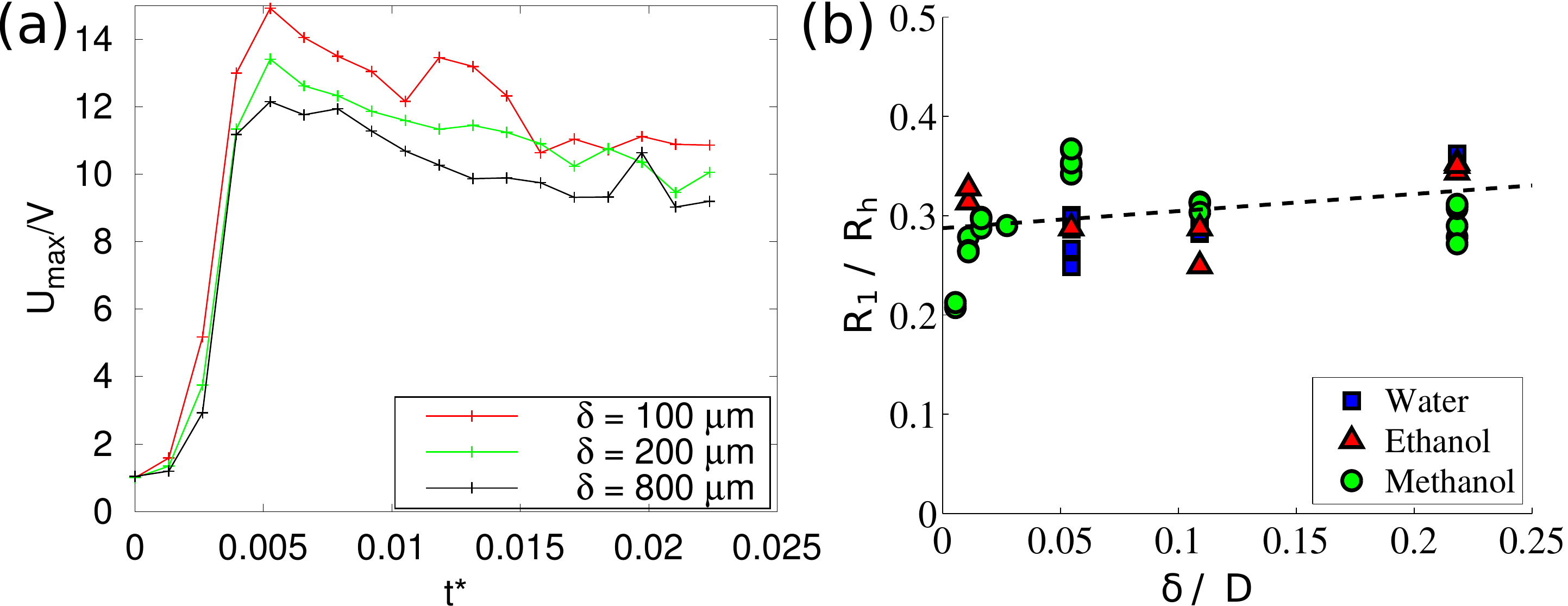}\vspace{0.12in}
\end{center}
\caption{(a) Maximum velocity U$_{max}$ in the liquid from the numerical simulations in Fig.~\ref{Fig_07},
for 3 different pool depths.
(b) Experimental observation of the first bubble ring entrapment radius
($Re=12\,900$, $\alpha = 0.80$) for different pool depths.
}
\label{Fig_07_2}
\end{figure}

In all three cases, we can see in the numerics that the central air disk punctures at the center
during its contraction into a central bubble, thus forming an air torus.
This is also observed in some of the experiments, as seen in Fig. \ref{Fig_08_2},
where an air torus is formed, which later contracts into one bubble.

\begin{figure}
\begin{center}
\includegraphics[width=0.88\linewidth]{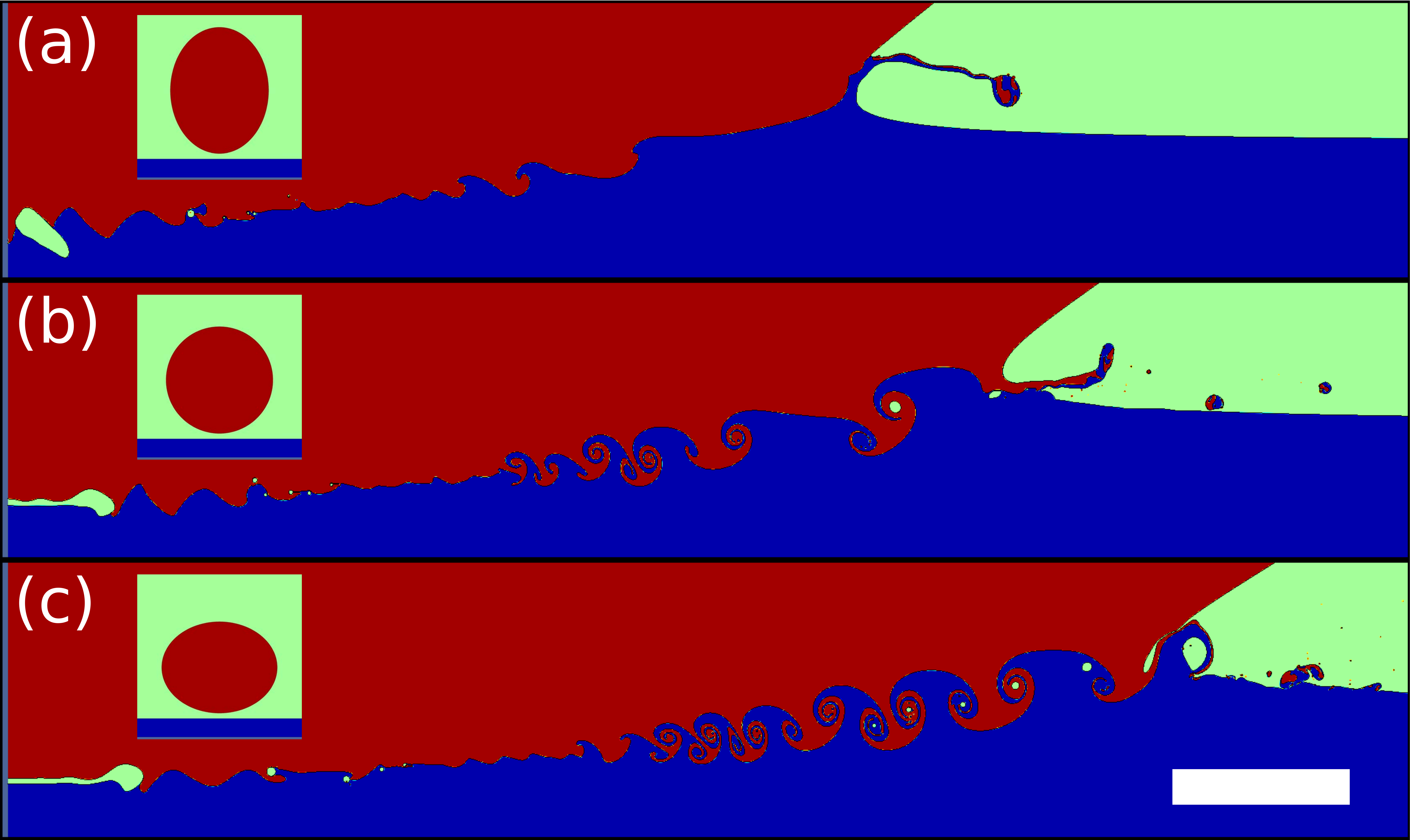}
\end{center}
\caption{Effect of drop shape on the early dynamics at t$^*$ = 0.0406,
for $Re = 12\,900$ and $\delta = 800\; \mu$m. The drop is modeled as an ellipsoid of revolution,
where $\alpha$ is (a) 1.29, (b) 1, (c) 0.79.
(a) and (c) correspond to the maximum horizontal deformations
of the fitting equation \eqref{eq:AvsT}, with $a = 0.162$,
keeping the same effective diameter.
The maximum and minimum level of refinement in the domain are respectively
5240 and 655 cells per drop diameter, and $D_{cut}$ = 3.1~$\mu$m.
The scale bar is 200 $\mu$m long.
}
\label{Fig_08}
\end{figure}

We observe experimentally that the most robust bubble rings are produced by a flat-bottom drop
in Fig.~\ref{Fig_02}(e) \& (f) and Fig.~\ref{Fig_04}(d-f) with fewest bubble rings in 7(e),
which is more spherical.
The largest number of rings is also produced by such oblate drops
(Fig.~\ref{Fig_02}(f) and Fig.~\ref{Fig_09}(b)).
This is consistent with the numerical results of Fig.~\ref{Fig_08},
showing a larger number of rings for the oblate drop.
It even suggests that a prolate drop could completely suppress
the bubble ring entrapments for the same effective diameter, as is shown in Fig.~\ref{Fig_08}(a).

\begin{figure}
\begin{center}
\includegraphics[width=\linewidth]{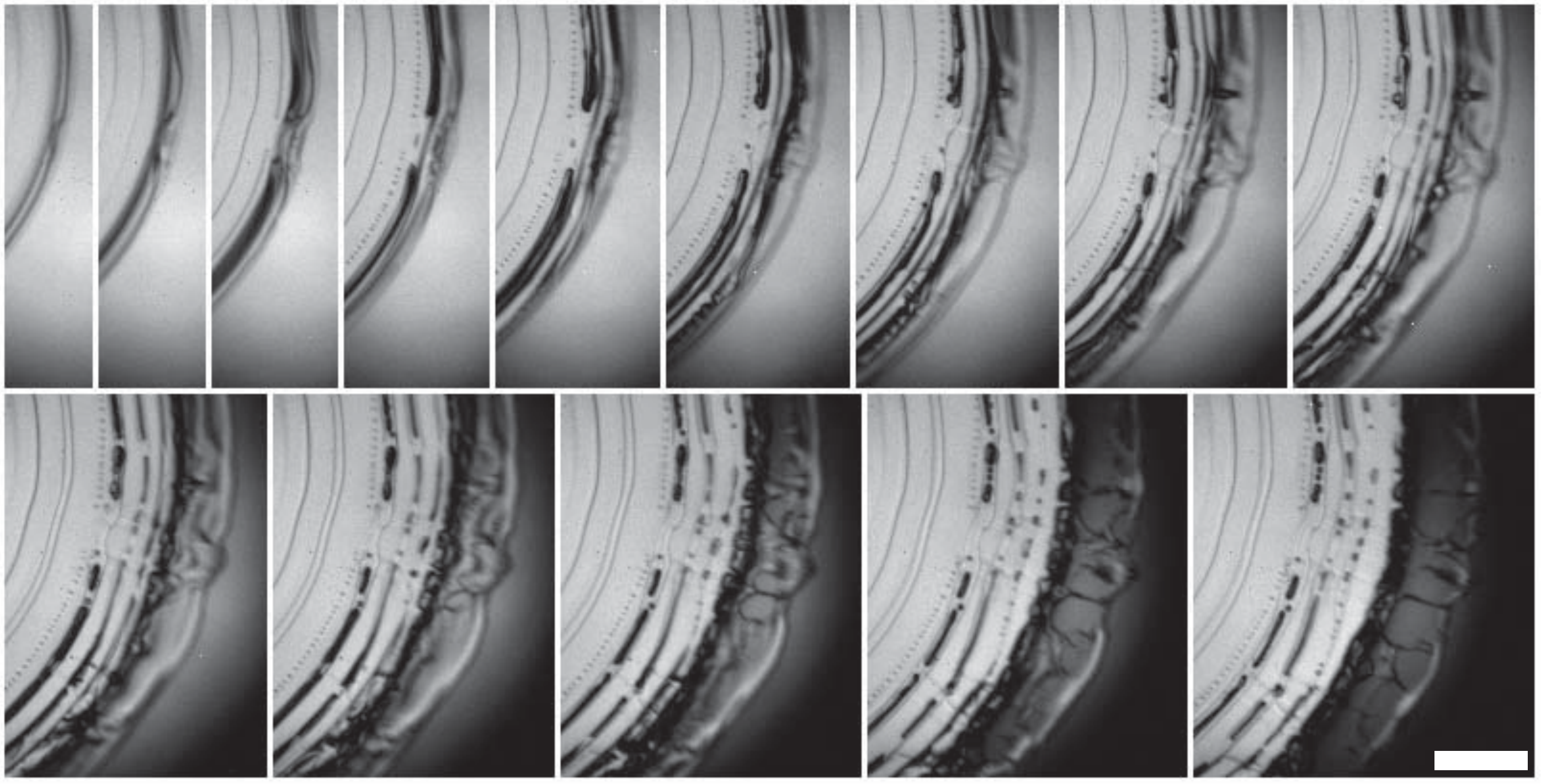}
\end{center}
\caption{Details of the ejecta sheet breakup;
for a water drop impacting a thin film of ethanol
($Re=11\,700$, $\alpha = 0.79$, $\delta = 75\; \mu$m).
The ejecta sheet starts to puncture on the second frame of the second row.
The growth of the holes leave tendrils connecting the neck to the liquid rim.
The scale bar is 200~$\mu$m long.
Frames are shown 2~$\mu$s apart.
}
\label{Fig_055}
\end{figure}

\begin{figure}
\begin{center}
\includegraphics[width=\linewidth]{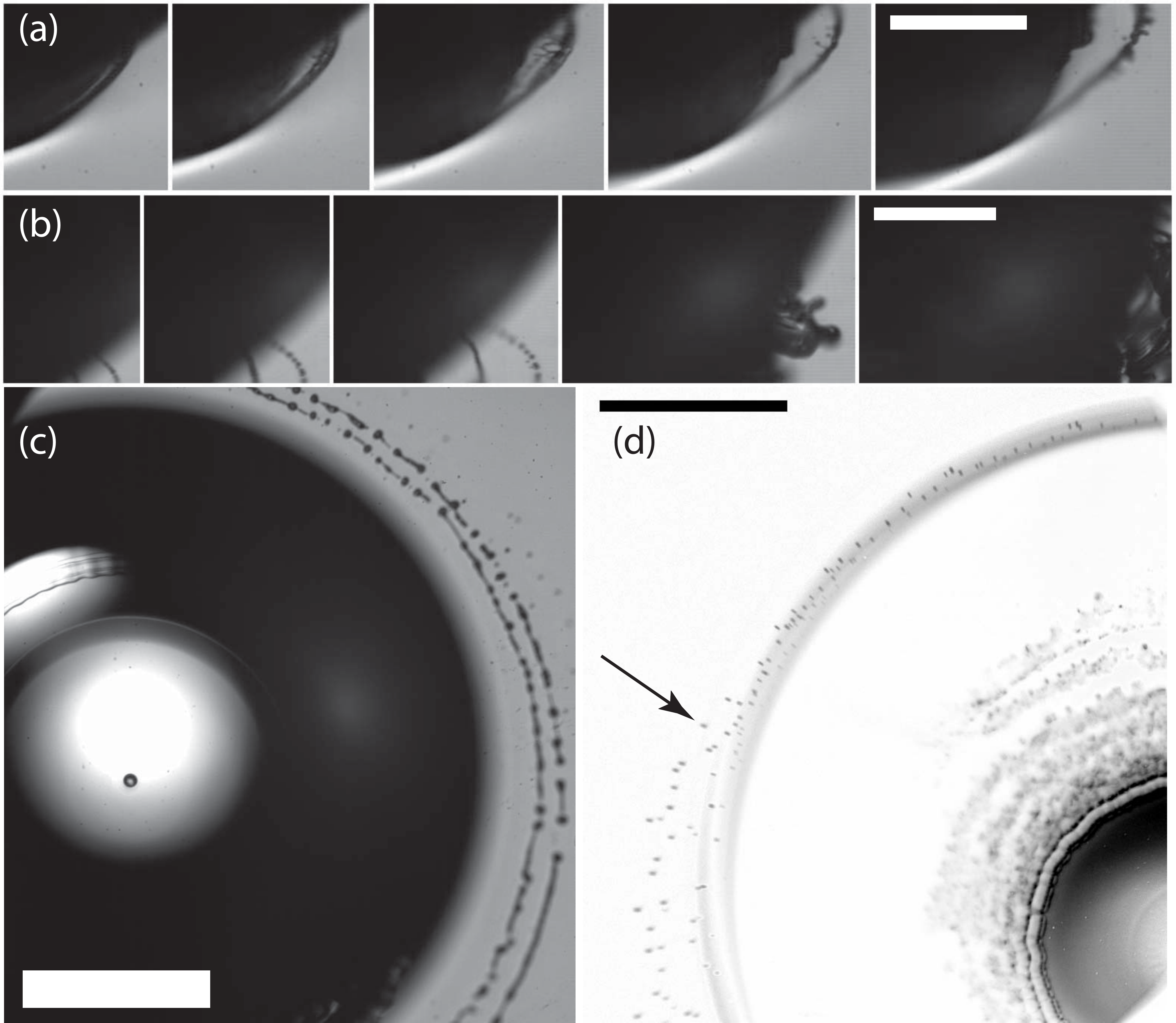}
\end{center}
\caption{Detachment of liquid toroids by breakup of the ejecta sheet.
Scale bars are 500 $\mu$m for (a,b) and 1 mm for (c,d).
All images have an exposure time of 1~$\mu$s.
(a) Side view of the liquid toroid detachment for a thin film of ethanol
($Re=12\,600$, $\alpha = 1.11$, $\delta = 125\; \mu$m).
The ejecta sheet punctures in the third frame, breaking into tendrils.
Those tendrils are slingshot in front of the liquid rim which has not yet broken,
as observed in the last frame.
The frames are shown 10~$\mu$s apart.
(b) Side view of two liquid toroid detachments for a thin film of methanol
($Re=12\,600$, $\alpha = 1.11$, $\delta = 125\; \mu$m).
The frames are shown 10~$\mu$s apart for the first 3 frames, and then 80~$\mu$s.
The splashing of liquid toroids is followed by the emergence of an irregular ejecta sheet,
as can be seen also on the larger bottom view of Fig.~\ref{Fig_10}.
(c) Larger bottom view of two consecutive rings of liquid droplets
and toroidal sections detaching for an ethanol pool
($Re=11\,600$, $\alpha = 0.86$, $\delta = 125\; \mu$m).
(d) First droplets emerging from underneath the drop in the same conditions as Fig. \ref{Fig_10}.
Image difference between two adjacent frames, to highlight the splashing droplets.  
Shown in inverted gray-scale.  The exposure is 1 $\mu$s long.
}
\label{Fig_056}
\end{figure}

\subsection{Edge breakup and splashing}
\label{sec:EdgeBreakup}
Numerical simulations have shown that the ejecta sheet can impact alternatively on the drop
and the pool during the vortex shedding.  The tip of the ejecta can thus detach into 
a liquid torus exiting the neck region at high speed \citep{Thoraval2012}.  
Such tori are highly unstable to Rayleigh instability and break rapidly into splashed micro-droplets.
However, this earliest splashing of micro-droplets by axisymmetric breakup of the ejecta sheet
had not been observed previously in experiments.

By looking carefully at Figs.~\ref{Fig_04}(d-f) \& \ref{Fig_10},
we can identify this early splashing by the breakup of the tip of the ejecta sheet
after the entrapment of a few bubble rings, as was suggested by the numerical simulations.
The liquid toroid in Fig.~\ref{Fig_04}(e) separates at $t^{*} \simeq$ 25 $\mu$s after the first contact,
and a velocity of 20.5~m/s, which corresponds to 7.1 times the impact velocity.
The tip velocity of the ejecta sheet in the numerical simulations at the same non-dimensional time,
is 6.2 for $Re = 12\,900$ (Fig.~\ref{Fig_08}(b)),
and 7.4 for $Re = 13\,700$ (Fig.~\ref{Fig_07}(b)),
which is in excellent agreement with the experimental observations.

Figure~\ref{Fig_055} shows more detail of the breakup of the edge at a slightly lower $Re$.
The ejecta sheet breaks via holes puncturing behind the rim.
The thicker ejecta rim is therefore left connected to the neck by liquid tendrils,
but subsequently becomes fully detached.

To remove the ambiguity of the bottom view, we have also looked at this early splashing
from the side above the free surface.
Figure~\ref{Fig_056}(a) confirms the bottom view images, showing the ejecta sheet emerging from the neck,
puncturing behind the rim and separating a liquid toroid from the neck.
Liquid tendril are also observed in the fourth frame, and are slingshot ahead of the rim,
creating the protrusions observed in the last frame,
similar to the last frame of the first row of Fig.~\ref{Fig_04}(e).
The slingshot of the broken ejecta sheet can also be observed in numerical simulation,
as in supplemental video of Fig.~\ref{Fig_08}(a),
and is similar to the slingshot mechanism described in \cite{Thoroddsen2011}.

Two consecutive liquid rings are observed in Fig.~\ref{Fig_056}(b).
It also shows the emergence of a greatly disturbed ejecta sheet after this early splashing.
A larger bottom view confirms this side view observations in Fig.~\ref{Fig_056}(c).
This mechanism, of a detachment of a thin torus of liquid,
explains the synchronized emergence of uniform sized micro-droplets observed ahead of
the main irregular ejecta sheet, see Fig.~\ref{Fig_056}(d) as well as Fig. 2(c) in Thoroddsen (2002).
The irregular sheet is clearly shown in Fig. 8(b).

The splashing of several liquid tori is consistent with numerical simulations showing that the ejecta sheet
can breakup during the successive impacts on the drop and the pool.
Supplemental video of Fig.~\ref{Fig_08}(a) shows such a case where the ejecta sheet breaks
first by climbing on the drop and then impacting on the pool, thus creating two consecutive liquid tori.

After the first ring entrapment, regular spanwise instabilities can appear in the ejecta sheet,
as is clearly seen in the second panel of Fig.~\ref{Fig_04}(d), as well as Fig.~\ref{Fig_04}(e) \& (f).
The fine azimuthal breakup when the ejecta bends and impacts onto a pool have been reported 
by \cite{Thoroddsen2011} (their Fig. 5) and may be of similar origin.
Furthermore, the early appearance of similar azimuthal instabilities have also been observed by \citet{Thoroddsen2012} 
in a free-surface cusp, which is formed during a drop impacting onto a solid surface.
Numerical simulations show that the ejecta sheet breaks when it impacts on the drop or the pool,
by stretching between the new connection and the faster rim.
This instability is therefore consistent with the impact of the ejecta sheet on the drop or the pool.
Similar breakup of a liquid sheet by stretching was also observed experimentally by \citet{Roisman2007} for spray impacts.

\subsection{Vortex shedding and rotation around bubble rings}
\label{sec:Rotation}
The difference in refractive index between the drop and the pool (see Table \ref{table01})
allows us to visualize vorticity structures inside the liquid.
As the coherent vortices bend and wrap up the interface between the two liquids,
a dark line can be observed at their edges with our back-light imaging setup.

Numerical simulations have shown that the first oscillations of the base of the ejecta sheet
have a smaller amplitude and do not entrap any bubble rings,
see \cite{Thoraval2012} (their Fig.~4(c,f,g)) and our Figs.~\ref{Fig_07} \& \ref{Fig_08}.
This is consistent with our experimental observations of Figs.~\ref{Fig_04}(e),
\ref{Fig_055}, \ref{Fig_03}(a) \& \ref{Fig_09},
where dark arcs form before the first bubble rings.
They show the shedding of vortices from the neck before the start of the bubble-ring entrapment.

\begin{figure}
\begin{center}
\includegraphics[width=\linewidth]{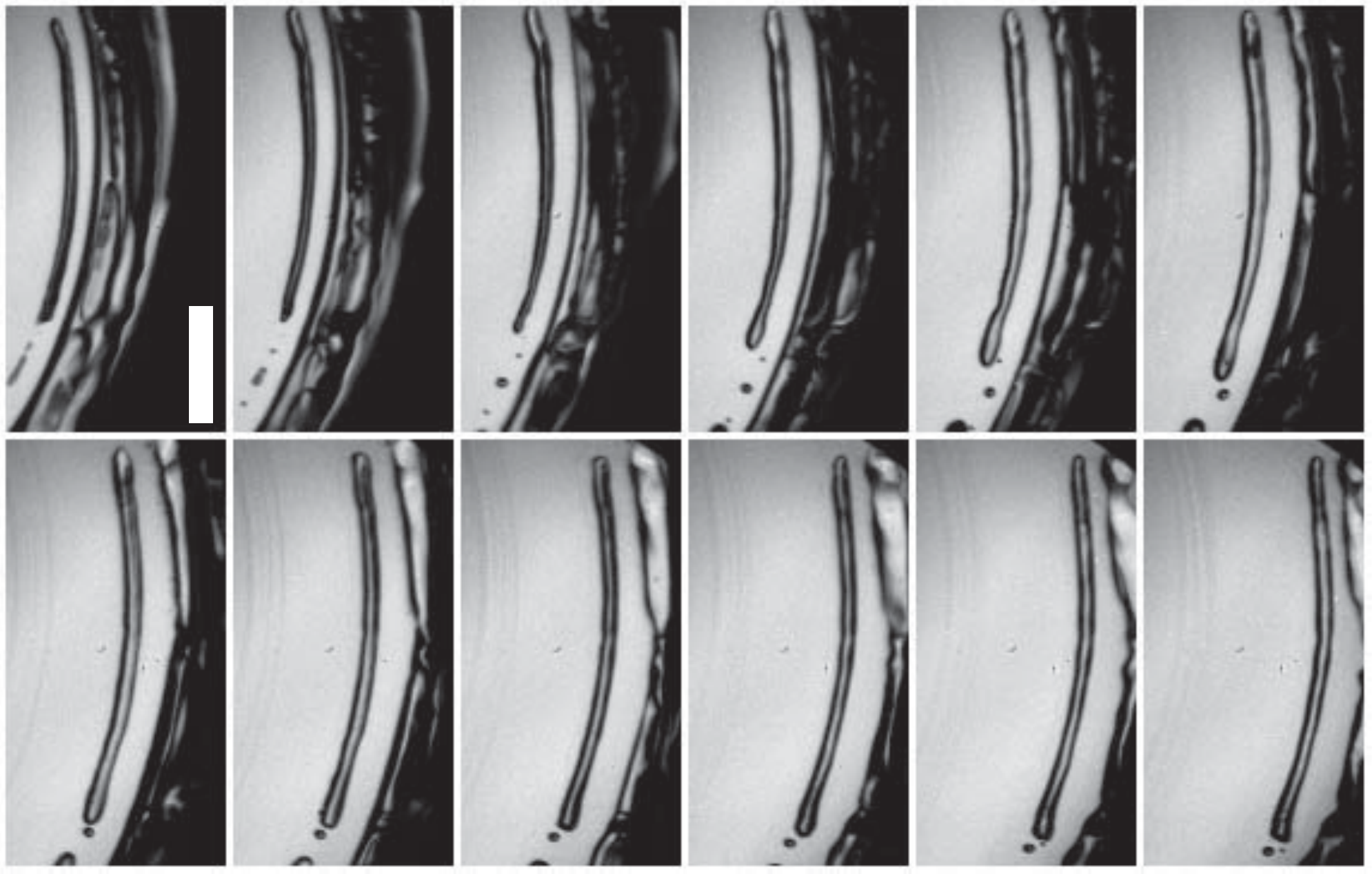}
\end{center}
\caption{Entrapment of a large air cylinder for a water drop impacting a thin film of methanol
($Re=11\,400$, $\alpha = 1.05$, $\delta = 50\; \mu$m).
The small bubble at the bottom right of the long cylinder make a complete rotation
around the cylinder between frames 4  and 12.
The frames are shown 4~$\mu$s apart.
The scale bar is 200~$\mu$m long.
}
\label{Fig_031}
\end{figure}

\begin{figure}
\begin{center}
\includegraphics[width=0.68\linewidth]{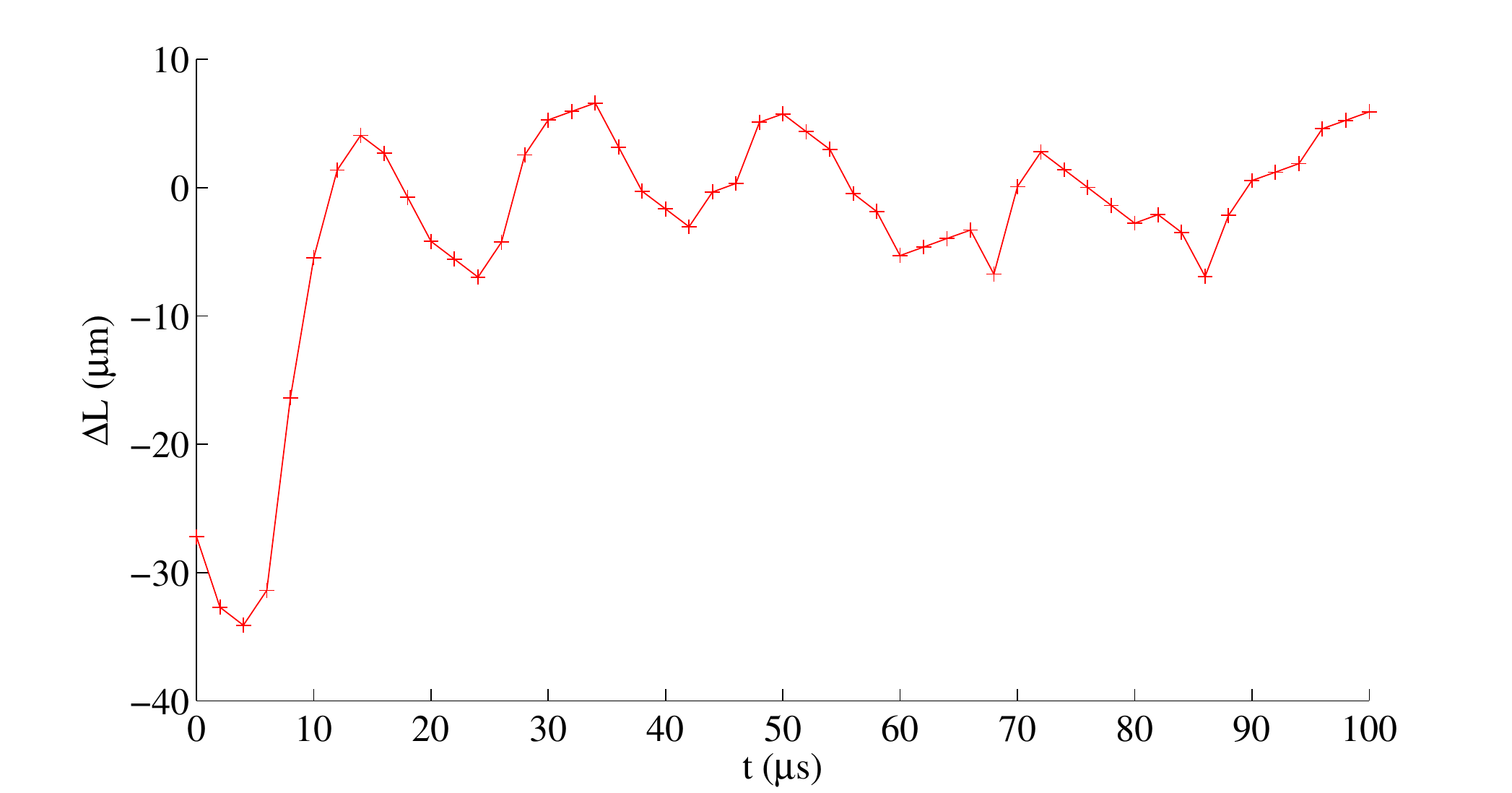}
\end{center}
\caption{Tracking of a 2~$\mu$m particle rotating around a vortex core for a drop impacting
on a pool of ethanol ($Re=11\,400$, $\alpha = 1.05$, $\delta = 125\; \mu$m).
$\Delta$L is the distance to the initial positions,
minus the mean radial translation component identified by a linear regression.
The period of rotation is here about 18 $\mu$s.
}
\label{Fig_032}
\end{figure}

\begin{figure}
\begin{center}
\includegraphics[width=1.0\linewidth]{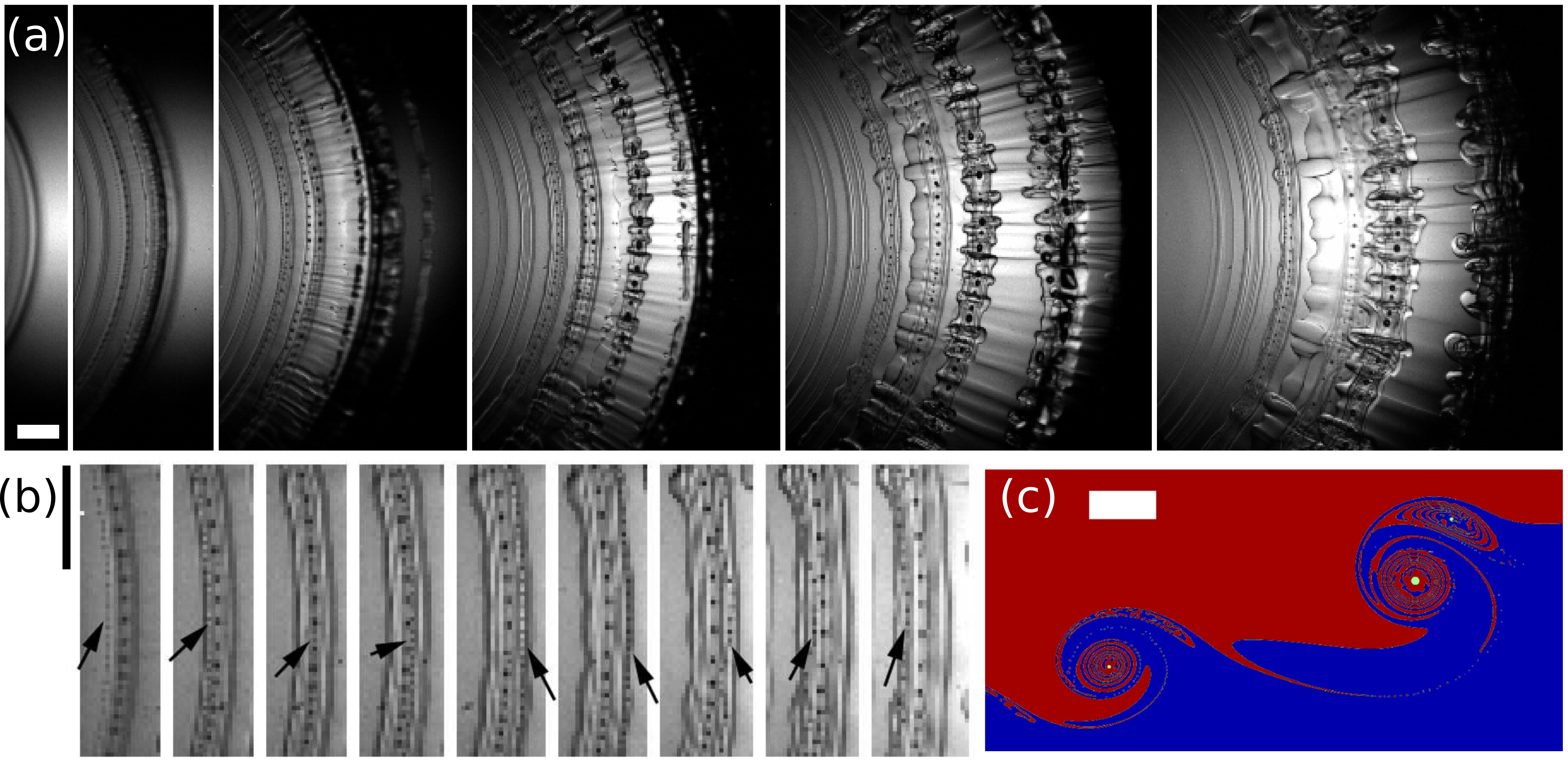}\vspace{0.15in}
\end{center}
\caption{(a) Bubble rings formed during the impact of a water drop onto a shallow pool of ethanol 
($Re=13\,300$, $\alpha = 0.98$, $\delta = 250\; \mu$m).
The frames are shown at 8, 20, 38, 66 \& 112 $\mu$s after the first one,
showing a total of 7 bubble rings.
(b) Close-up view of a line of fine bubbles which circulate around another line of slightly larger bubbles,
from the video in (a). Total duration of these frames is 104 $\mu$s.
See also supplemental videos.
(c) Numerical simulations of a water drop impacting on a thin film of the same liquid
($Re=13\,800$, $\alpha = 1$, $\delta = 800\; \mu$m).
This focused view of the interface at t* = 0.481 shows
the rotation of a pair of bubble tori at the core of adjacent vortex rings.
The scale bars are all 100 $\mu$m long.
}
\label{Fig_03}
\end{figure}

As the neck moves outwards radially, the angle between the pool and the drop becomes larger,
and bubbles rings are entrapped, as observed above.
Numerical simulations have shown that these rings are often entrapped in an alternating way
at the top and the bottom of the ejecta sheet.
At the same time, they shed vortices of alternating sign in the liquid.
Bubble rings are therefore entrapped at the core of vortex rings.
Dark lines are indeed observed experimentally around the bubbles, supporting this shedding of vorticity scenario.
The rotation is also made apparent by the dynamics of a micro-bubble rotating around
a larger bubble cylinder in Fig.~\ref{Fig_031}.
We can obtain an estimate of the rotation speed in the vortex rings by tracking a 2~$\mu$m particle
seeded into the pool liquid (Fig.~\ref{Fig_032}), giving a rotation period of 18~$\mu$s in this example.

The dynamics of the air entrapped in this vortex street is also affected by the rotation around it.
One can expect the rotation to delay the capillary breakup of the air cylinders
\citep{Rosenthal1962, AshmoreStone2004, EggersVillermaux2008}.
This is indeed evident in Fig.~\ref{Fig_031}, where the cylinder of diameter about 29~$\mu$m
can be observed for more than 94~$\mu$s before breaking,
corresponding to $ t/\tau_{\sigma} \simeq$ 7 based on the methanol properties.
The air cylinder also elongates by about 25\% between the first and the last frame, consistent with the theory
that it resides inside a vortex.
The relative motion of the bottom tip of the air cylinder and the closest micro-bubble below
shows that the stretching due to the radial motion cannot account for this elongation.
Even smaller cylinders can be stabilized, as observed in Fig.~\ref{Fig_06}(c).
Two air tori are formed next to each other, with similar diameters $d_{tor}\simeq 8 \, \mu$m (see arrows in the figure),
but break up at very different times from formation, of $ t/\tau_{\sigma} \simeq$ 4 and 12,
based on the liquid properties of the methanol.
These delayed breakups show the strong stabilization effect of the circulation
around these air tubes.
The breakup wavelength is also larger, as the theory suggests \citep{AshmoreStone2004}.

The row of vortices shed in the liquid can also interact with adjacent ones.
In some realizations two closely entrapped air tori rotate around each other, with the line of small bubbles
rotating around the bigger one. Figure~\ref{Fig_03}(b) shows such a sequence,
where we track one rotation, which takes $T= 104\; \mu$s.
Numerical simulations show that vortices of different strength are created
at the top and bottom of the ejecta sheet \citep{Thoraval2012}.
The longer time evolution shows that this difference can make the bubble rings
rotate around each other while translating (Fig.~\ref{Fig_03}(c) and supplementary videos).
These dynamics are consistent with the experimental observation of the rotation around bubble rings.

Some of the bubble tracks simply translate past each other during their radial motions,
for example visible the videos accompanying Figs. 19(a) \& 21(a).
The bubbles are initially sitting at slightly different depths,
and this difference in vertical location is amplified by the vorticity, as shown above.
Moreover, as the bubble tori break into bubbles, their vertical width will slightly increase,
sampling larger mean shear.
These relative translations of bubbles
could therefore result from a vertical mean gradient of horizontal velocity within the pool depth.

\begin{figure}
\begin{center}
\includegraphics[width=0.82\linewidth]{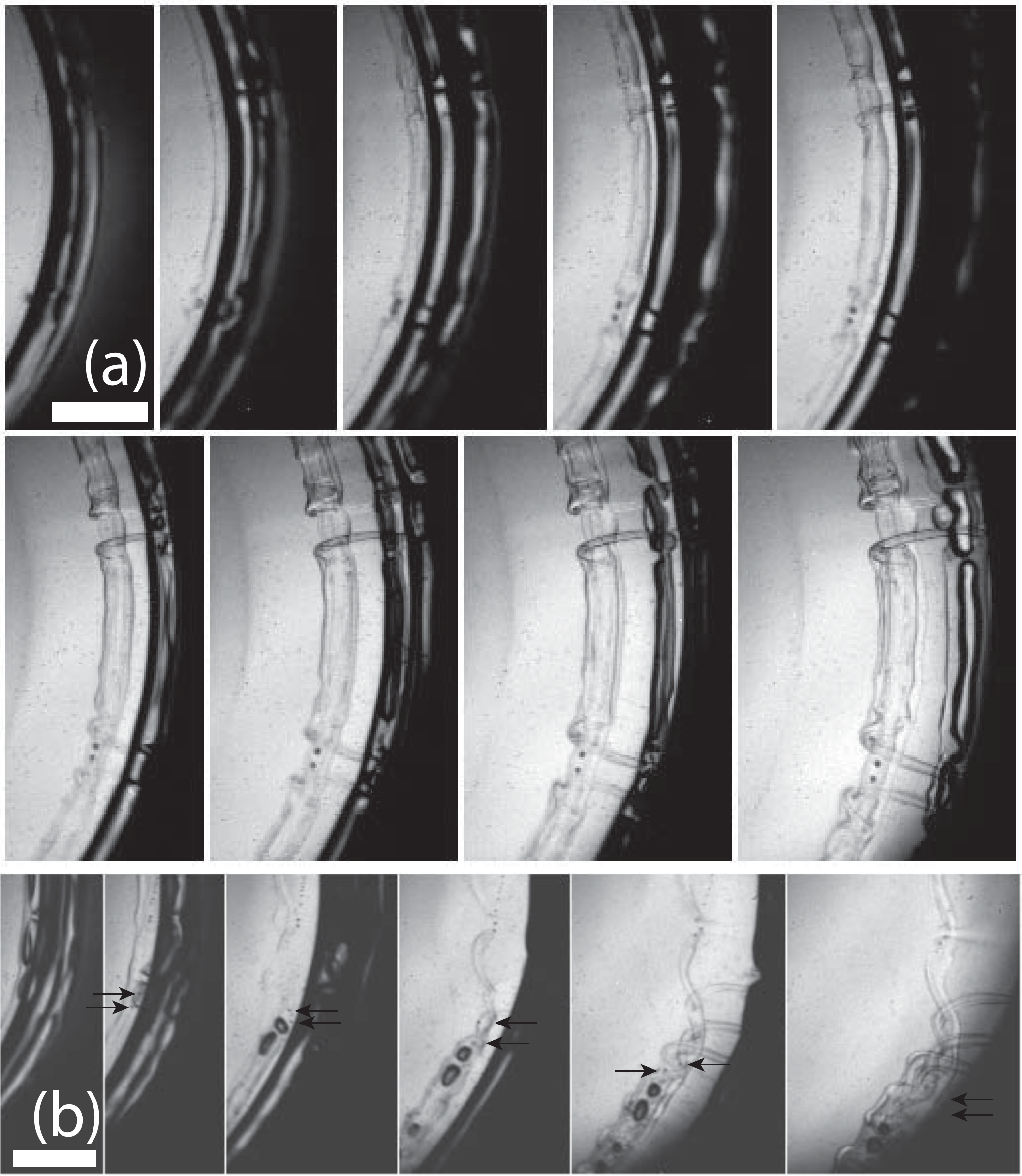}
\end{center}
\caption{(a) Roll-up of isolated streamwise vortex pairs
for a water drop impacting a film of ethanol seeded with 2~$\mu$m particles
($Re=11\,400$, $\alpha = 1.05$, $\delta = 125\; \mu$m).
The first images show that the vortex pairs are starting at the same azimuthal location
as local disturbances in the front.
The bottom vortex pair first entraps a bubble, later splitting in two.
Frames are shown 3, 6, 9, 12, 17, 23, 27, 35~$\mu$s after the first one.
(b) Roll-up and sideways motion of streamwise vortices
for a water drop impacting a film of ethanol seeded with 2~$\mu$m particles
($Re=11\,400$, $\alpha = 1.05$, $\delta = 125\; \mu$m).
Frames are shown 3, 10, 18, 32 \& 47~$\mu$s after the first one.
The scale bars are 200~$\mu$m long.
}
\label{Fig_12}
\end{figure}

\subsection{Three-dimensional instabilities}
\label{sec:Tangle}

The axisymmetry of the impact is rapidly broken by different instabilities
(see Figs.~\ref{Fig_02} \& \ref{Fig_04}).
For a water pool, we have observed undulations develop in the neck as soon as an ejecta forms (Fig.~\ref{Fig_02}).
This leads to entrapment of isolated bubbles (Fig.~\ref{Fig_02} \& \ref{Fig_022})
and creates less regular bubble rings compared to ethanol or methanol pools.
Even in those lower surface tension liquids, the most regular rings appear at smaller entrapment radii,
where the ejecta sheet has not broken yet and remains axisymmetric.

We have already suggested in \S\ref{sec:EdgeBreakup} that some of the azimuthal instabilities
come from the breakup of the ejecta sheet, when it bends and touches the drop or pool surfaces.
Overturning gravity waves also rebound and destabilize underlying vortices, 
but at a much larger scales than herein, see \cite{Watanabe2005}.
Figure~\ref{Fig_055} clearly demonstrates the effect of isolated neck disturbances on the bubble entrapment.
A small perturbation is visible in the first frame and develops in time.
The two first bubble rings which form in the following frames are broken at the same azimuthal location
and the ejecta ruptures there first.

\begin{figure}
\begin{center}
\includegraphics[width=1.0\linewidth]{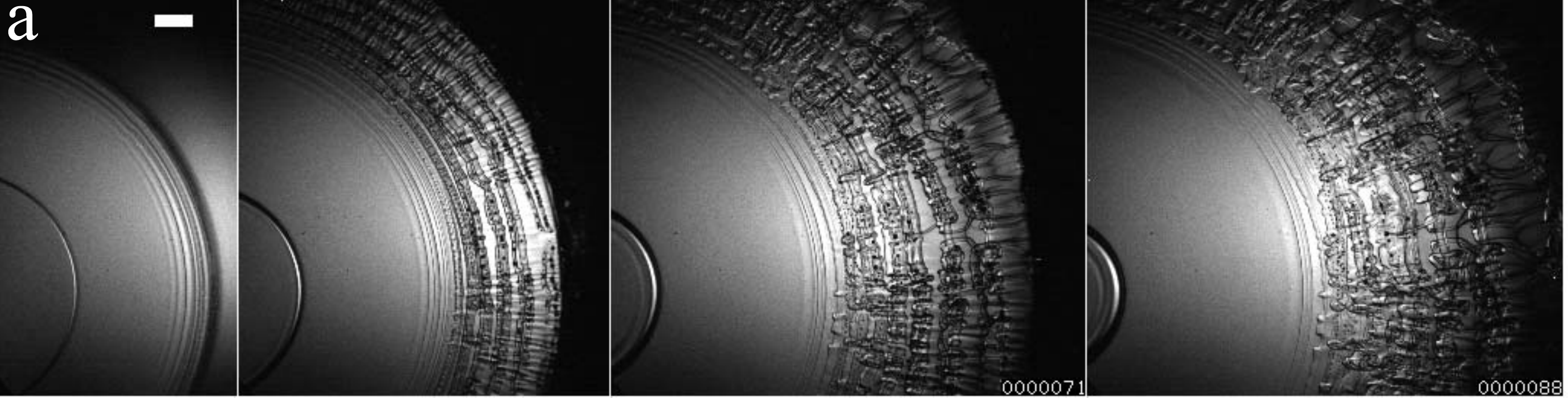}\vspace{0.15in}
\includegraphics[width=1.0\linewidth]{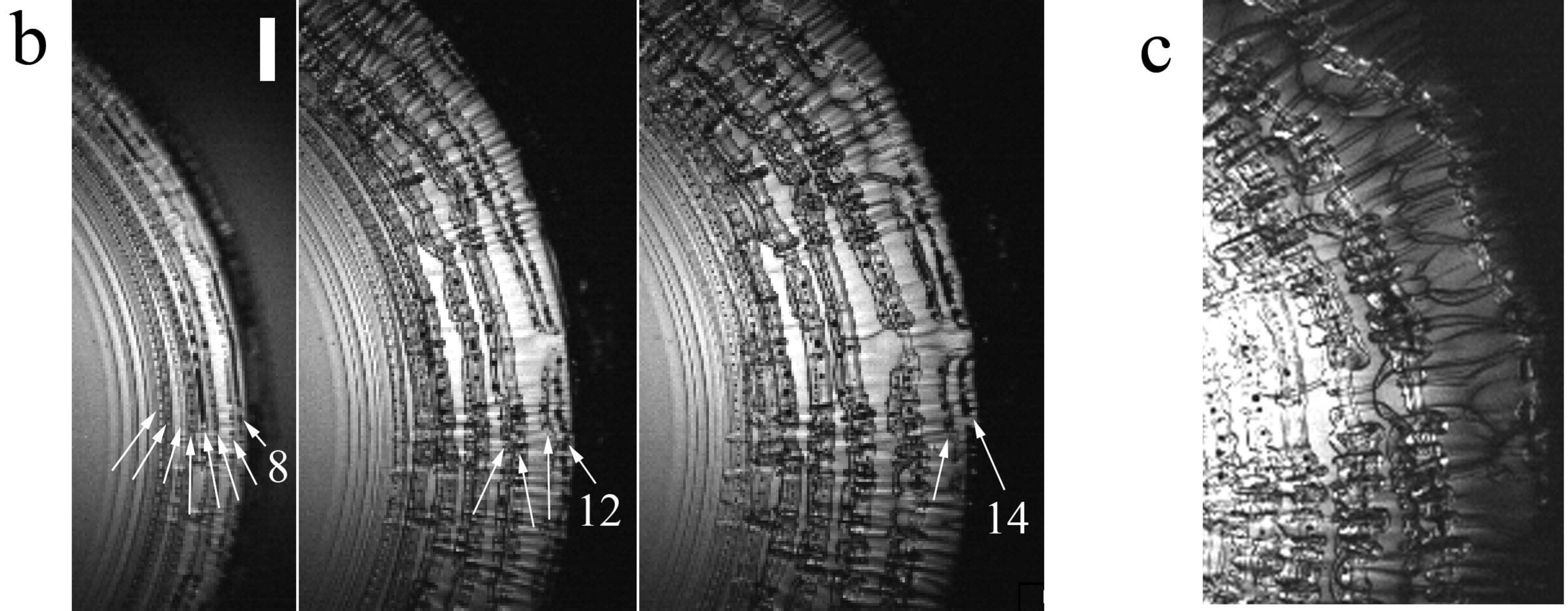}\vspace{0.15in}
\end{center}
\caption{Bubble rings and vortex entanglement ($Re=12\,900$, $\alpha = 0.80$).
(a) Frames shown are 8, 22, 41 \& 58 $\mu$s from first contact on ethanol ($\delta = 250\; \mu$m).
(b) Careful examination of the fine bubbles, show 14 separate bubble rings,
which are pointed out by the arrows.
(c) Closeup of the vortex tangles at $t=58\; \mu$s.
The scale bars are 100 $\mu$m long.
See also supplementary video.
}
\label{Fig_09}
\end{figure}

Three-dimensional instabilities also develop inside the liquid,
and are made apparent by the difference in refractive index.
Radial lines are visible in Figs.~\ref{Fig_04}(e), \ref{Fig_10},
\ref{Fig_03}(a), \ref{Fig_12} \& \ref{Fig_09}.
They show the formation of streamwise vortices between the primary spanwise vortex rings,
which often reach to the free surface in the neck.
For the lower $Re$ cases (Figs.~\ref{Fig_12}),
isolated streamwise vortices are observed.
In Fig.~\ref{Fig_12}(a), they appear in pairs, at the same location as a front perturbation.
The lower one in the image starts at an isolated location on the side of a vortex ring,
where it entraps one micro-bubble.
Two lines are then visible on each side of this initial entrapment and extend up to the front.
This suggests that the vortex pairs arises from the same vortex loop,
rolling-up around the vortex ring.
The connection between the two streamwise vortex lines should then form
a vortex loop with one section in the azimuthal direction, near its origin around the bubble.
The presence of a strong vorticity around this bubble
is demonstrated by its breakup into two smaller bubbles (in the 4th panel of Fig.~\ref{Fig_12}(a)).
Moreover, the later dynamics shows the secondary streamwise vortex tubes roll up 
around the primary spanwise vortex ring.
Figure~\ref{Fig_12}(b) shows a similar case where streamwise vortices roll-up around a vortex ring.
This roll-up can be identified by following two micro-bubbles at the core of the vortices
(black arrows).

Interesting parallels can be made with similar three-dimensional instabilities occurring
in the wake of a cylinder \citep{Williamson1996}.
The local Reynolds number at the base of the ejecta sheet $Re_{b}$ will be affected by both
the radial velocity of the neck and the velocity within the liquid.
$Re_{b}$ can be increased both by increasing the impact velocity
or by using a more oblate drop.
Indeed, as the impact $Re$ increases, the ejecta sheet emerges from
a faster moving neck \citep{JosserandZaleski2003,Thoraval2012},
and at a higher velocity \citep{Thoroddsen2002, JosserandZaleski2003},
thus leading to higher $Re_{b}$.
A flat bottom drop also geometrically generates a faster moving neck,
and produces larger velocities in the liquid, see Fig.~\ref{Fig_07_2}(a).
Both effects lead to a larger concentration of streamwise vortices
(Figs.~\ref{Fig_04}(e), \ref{Fig_10}, \ref{Fig_03}(a), \ref{Fig_12} \& \ref{Fig_09}).
This is similar to the onset of three-dimensional instabilities of the vortex street behind a circular cylinder
\citep{Williamson1988, Williamson1996},
where finer streamwise vortices and a smaller spanwise wavelength are observed at higher $Re$.
Similar vortex loops are also observed in both cases, as described above.
Moreover, rapid motion of the bubbles in the spanwise direction along the vortices
is observed in our experiments, see Figs.\ref{Fig_02}(f) and \ref{Fig_04}(f).
Similar lateral motion was also observed behind ``vortex dislocations''
in the wake of a cylinder \citep{Williamson1992, Williamson1996}.

\section{Conclusions}
\label{sec:Conclusions}

Observations from below the impacting drop have herein demonstrated that the mechanism suggested
in \citet{Thoraval2012} does indeed entrap micron sized air tori.
The oscillations of the ejecta sheet can thereby entrap a row of bubble rings.
The vorticity entering the liquid during those oscillations and bubble entrapments
is then destabilized into complex 3D structures.
The combination of azimuthal instabilities and vertical oscillations produces
a large range of new bubble entrapment scenarios.

Besides imaging the formation and breakup of partial bubble rings, 
of equal significance is our observation that the outer neck is unstable to
azimuthal undulations, for water on water impacts at even very moderate Reynolds numbers (Figs. 4(d) \& 5).
This poses a challenge to theoretical and numerical studies, which invariably assume axisymmetry.

We note that the bubble rings observed herein differ in fundamental ways 
from the Oguz-Prosperetti rings \citep{OguzProsperetti1989},
as the base of the ejecta is not driven by surface tension, but rather by the impact pressure.
This high localized pressure is indeed the mechanism responsible for driving out the ejecta sheet.

However, the details of the air entrapment and its dependence on the impact conditions is still not clear.
Large scope exists for further work, as the present study perhaps raises as many questions as it answers.
What role do Marangoni or Rayleigh-Taylor play in the two-liquid dynamics?  
The large parameter space of other liquids needs to be studied.
Even for the same liquids, in the drop and pool, the interplay between $Re$, $We$ and $\alpha$
which allows entrapment, or preserves an extended axisymmetric ejecta, remains to be determined.
The three-dimensional instabilities of the vortex street also need to be studied in more detail
and compared to the well-known instabilities of the cylinder wake and the shear layer.
However, in the vortex street observed here, the vortices are shed from a deformable free-surface,
adding to the complexity of the analysis.
The influence of the shed vorticity on the dynamics of the neck,
both vertically and in the azimuthal direction, should also be added
in the stability analysis of splashing.

\end{document}